\documentclass[aps,prl,reprint,superscriptaddress]{revtex4-2}

\usepackage[utf8]{inputenc}
\usepackage{amsmath}
\usepackage{amssymb}
\usepackage{comment}
\usepackage{float}
\usepackage{caption}
\usepackage{xcolor}
\definecolor{dgreen}{rgb}{0,0.5,0}
\definecolor{dpink}{rgb}{1,0.3,0.3}
\definecolor{darkblue}{rgb}{0,0,0.6}
\definecolor{purple}{rgb}{0.4,.2,0.7}
\usepackage[margin = 2.5cm]{geometry}
\usepackage{graphicx}
\usepackage{tikz}
\usetikzlibrary{decorations.markings}
\tikzset{
midarr/.style={postaction={decorate},decoration={markings,mark=at position 0.55 with {\arrow{>}}}},
midcirc/.style={
    postaction={decorate},
    decoration={markings, mark=at position 0.55 with {\draw[fill=gray] circle (0.15);}}
  },
bprop/.style={thick,dashed,midarr},
yline/.style={thick,midarr},
ylineR/.style={thick,midarr},
ydouble/.style={thick,double,double distance=1.4pt,midarr},
Xdouble/.style={thick,double,double distance=1.4pt,midcirc}
}
\usepackage[hyperfootnotes = false, colorlinks = true, linkcolor = blue, citecolor = purple]{hyperref}
\usepackage{subcaption}
\usepackage{multirow}
\usepackage{booktabs}

\usepackage{ytableau}

\usepackage{fancyhdr}
\parskip=\medskipamount
\newgeometry{margin=2cm}

\makeatletter
\def\section{%
  \@startsection
    {section}%
    {1}%
    {\z@}%
    {0.6cm \@plus1ex \@minus .2ex}%
    {0.35cm}%
    {%
      \normalfont\small\bfseries
      \centering
    }%
}%
\def\subsection{%
  \@startsection
    {subsection}%
    {2}%
    {\z@}%
    {0.6cm \@plus1ex \@minus .2ex}%
    {0.35cm}%
    {%
     \normalfont\small\bfseries
     \centering
    }%
}%
\makeatother

	\newcommand{\RA}{\Rightarrow}

	\newcommand{\hf}{\frac{1}{2}}
	
	\newcommand{\qrt}{\frac{1}{4}}

    \newcommand{\ym}{{\mathrm{YM}}}
	\usepackage{physics}
	\renewcommand{\ev}[1]{\langle #1 \rangle} %
	
	\def\i{\mathrm{i}}

\def\co{\mathcal{O}}

\begin{document}

\title{High-Precision Bootstrap of Multimatrix Quantum Mechanics}

\author{Henry W. Lin}
\email[]{hwlin@princeton.edu}
\affiliation{Leinweber Institute for Theoretical Physics, Stanford University, Stanford, California 94305, USA}
\affiliation{Jadwin Hall, Princeton University, Princeton, New Jersey 08540, USA}

\author{Zechuan Zheng}
\email[]{zechuan.zheng.phy@gmail.com}
\affiliation{Perimeter Institute for Theoretical Physics, Waterloo, Ontario N2L 2Y5, Canada}
\affiliation{Laboratoire de Physique de l\textquoteright Ecole Normale Sup\'erieure, ENS, Universit\'e
PSL,  \\
CNRS, Sorbonne Universit\'e, Universit\'e Paris Cit\'e, F-75005 Paris, France}

\date{\today}

\begin{abstract}
We consider matrix quantum mechanics with multiple bosonic matrices, including those obtained from dimensional reduction of Yang-Mills theories. Using the matrix bootstrap, we study simple observables like $\langle \mathop{\tr} X^2 \rangle$ in the confining phase of the theory in the infinite $N$ limit. Exploiting the symmetries of these models and applying nonlinear relaxation, we impose constraints that include traces of words of length up to $14$. Our results yield rigorous bounds on the large-$N$ ground-state dynamics, along with estimates of selected low-order observables to eight significant digits.
\end{abstract}


\maketitle

\section{Introduction} 
Solving large $N$ gauge theories has been a long-held dream of modern physics.
An important tool in this pursuit has been Euclidean Monte Carlo lattice computations. However, this approach requires extrapolation to large $N$ where computation becomes increasingly expensive, although physics simplifies in this limit. Another nonperturbative approach that has comparatively been less developed is the matrix bootstrap \cite{Anderson:2016rcw, Lin:2020mme, Han:2020bkb, Kazakov:2021lel} which works directly in the infinite $N$, 't Hooft limit by imposing large $N$ factorization \footnote{Alternatively, at finite $N$ one may impose trace relations \cite{Kazakov:2024ool}.}.

In this work, we continue developing this approach by performing a high-precision bootstrap of large $N$ bosonic matrix quantum mechanics. These include models arising from dimensional reduction of Yang-Mills (YM) theories in $D+1$ dimensions. Such $0+1$-dimensional models share some qualitative aspects with their higher-dimensional counterparts in compact spaces \cite{Aharony:2003sx}, with a similar phase diagram. More precisely, it is believed that, at large $N$, these quantum mechanical theories exhibit two continuous phase transitions.  
For low temperatures $T < T_{c,1}$, the theory is in a confining phase characterized by a $\sim O(N^0)$ free energy. At the first critical temperature, $T_{c,1}$, the theory undergoes a second-order Hagedorn transition to a deconfined phase, where the free energy scales as $O(N^2)$. Another third-order Gross-Witten-Wadia transition~\cite{Gross:1980he, Wadia:1980cp} occurs at a temperature $T_{c,2} > T_{c,1}$. Both transitions occur at temperatures of order one in units of the 't Hooft coupling~\cite{Aharony:2004ig, Kawahara:2007fn, Mandal:2009vz}.

A key improvement over previous bootstrap studies %
is the use of nonlinear relaxation techniques to deal with the nonlinear system of equations that arise after imposing a large $N$ factorization \cite{Kazakov:2021lel}. So far, this method has been implemented in the path integral context \cite{Kazakov:2021lel, Kazakov:2022xuh}; in this work, we implement it in the Hamiltonian formalism. We also apply the ground state positivity condition \eqref{gsp}, which dramatically boosts the precision of the bootstrap. Finally, we systematically leverage the O($D$) symmetry of these models which was crucial in allowing us to go up to level 14.

This work may also be viewed as a natural continuation of~\cite{LinZheng1}.  
The bosonic models considered here are somewhat simpler to bootstrap than the supersymmetric Banks-Fischler-Shenker-Susskind (BFSS) model studied in~\cite{LinZheng1}, mainly because the group-theoretic structure is less intricate. The computational methods developed in this work should inform future efforts, including our planned high-precision bootstrap analysis of the BFSS matrix quantum mechanics~\cite{LinZheng2}.

Our motivation to push the precision frontier is not merely precision for the sake of precision. 
In the supersymmetric BFSS case, a high-precision measurement of the thermodynamics and/or correlation functions would allow one to ``experimentally'' measure higher derivative corrections to Type IIa supergravity \cite{Kabat:2000zv, Anagnostopoulos:2007fw, Hanada:2008ez, Catterall:2008yz, Filev:2015hia, Kadoh:2015mka, Berkowitz:2016jlq, Berkowitz:2018qhn, Pateloudis:2022ijr,Hoppe:1999xg}.
These corrections are currently unknown using world sheet methods. So, the high-precision bootstrap could teach us about quantum gravity. In a similar spirit, lattice Monte Carlo studies for large $N$ Yang-Mills theory (see \cite{Lucini:2012gg} for a review and e.g. \cite{Caselle:2024zoh, Athenodorou:2024loq} for recent progress) can teach us about the effective string. We view our work as a starting point for a bootstrap study for higher-dimensional Yang-Mills theory in the Hamiltonian \cite{Kogut:1974ag} lattice gauge theory formalism \footnote{See \cite{Anderson:2016rcw, Kazakov:2022xuh, Kazakov:2024ool, Li:2024wrd, Guo:2025fii} for some bootstrap approaches to the Euclidean path integral approach to pure Yang-Mills theory.}.

Several appendices complement this work.
In Appendices~\ref{app: details}--\ref{app: adjoint}, we provide details on the numerical implementation, the relaxation procedure, some results on large $D$ expansion, and further discussion of the adjoint gap.
In Appendix~\ref{app: example}, we include a pedagogical example that illustrates our bootstrap framework in a simplified setting. We also describe a specific method for performing the irreducible decomposition of operators in Appendix~\ref{app: irrep} and summarize the technical steps involved in the large-$D$ analysis in Appendix~\ref{app: largeD}.

\section{Bootstrap ingredients \label{sec: ingredients}}
\subsection{The models}
We consider systems that consist of $D$ bosonic matrices, satisfying the commutation relations $[(X_I)_{ij},(P_J)_{kl}] = \i \, \delta_{il} \delta_{jk} \delta_{IJ}$. Here $I,J$ run from 1 to $D$. These matrices are traceless and Hermitian and transform in the adjoint of SU($N$). The Hamiltonian is
\begin{small}
\begin{align}
        H =& \hf \sum_{I=1}^D  \Tr P_I P_I + M^2  \Tr X_I X_I  \nonumber \\
        &- \frac{g^2_\ym }{4} \sum_{I,J} ^D \Tr [X_I , X_J]^2 
\end{align}
\end{small}
\noindent
The ground state of this model is gauge invariant $C \ket{\Omega}=0$, where the SU($N$) gauge generators are $C = \sum_{I=1}^D ( -\i [X_I, P_I]   - N\mathbf{1})$. The model possesses an $\mathrm{O}(D)$ global symmetry, under which the fields $X_I$ and $P_I$ transform in the fundamental representation.  
The case with $M^2 = 0$ is of particular interest, as it arises from the dimensional reduction of pure Yang--Mills theory in $D+1$ dimensions.  
(For certain values of $D$, the massless theory may also be obtained by reducing supersymmetric Yang--Mills theories on tori $T^d$ with anti-periodic boundary conditions for the fermions; see, for example,~\cite{Aharony:2004ig}.)  

For the massless model, we may set $g^2_{\mathrm{YM}} = 1$ without loss of generality by performing a canonical transformation together with a rescaling of the Hamiltonian.  
For the massive model, the dynamics are governed by the dimensionless effective coupling
$\lambda_{\mathrm{eff}} = g^2_{\mathrm{YM}}\, N/M^3$.

For the $D=2$ version, it is convenient to view $\mathrm{SO}(2)$ as $\mathrm{U}(1)$ and work with complex operators 
\begin{align}
    Z &=\frac{1}{\sqrt{2}} (X_1 + \i X_2), \quad \bar{Z} = \frac{1}{\sqrt{2}} (X_1 - \i X_2),
\end{align}
and similarly  $P = \,\frac{1}{\sqrt{2}}( P_1 + \i P_2)$, 
$\bar{P} =\frac{1}{\sqrt{2}} \, (P_1 - \i P_2).$
Then the $\mathrm{U}(1)$ charge of an operator is simply the difference between (number of $Z$'s and $P$'s) $-$ (number of $\bar{Z}$'s and $\bar{P}$'s). The additional $\mathrm{O}(2)$ reflection symmetry sends $Z \leftrightarrow  \bar{Z}$.
With this notation, 
 $   H=  \tr P \bar{P} + M^2 \tr Z \bar{Z} + \hf \tr [Z,\bar{Z}]^2$
and $C =  -\i [Z, \bar{P}] -\i [\bar{Z},P]  - 2$. 

Because of large-$N$ factorization, we only need to consider single-trace operators. The operators are ``words" composed of ``indexed letters" $X_I$ and $P_J$ with indices fully contracted. Since  $\Tr X_I P_I = \frac{D}{2} \i N^2$, we perform the following shift~\footnote{This is equivalent to introducing a formal $\hbar = 1/N$ so that $[X_{ij},P_{kl}] \propto \i \hbar $.} so that expectation values of single-trace operators are of order one in the large $N$ limit:
\begin{equation}\label{largeNscaling}
    \Tr \rightarrow \tr = \frac{1}{N}\Tr, \;
    X_I\rightarrow\frac{1}{\sqrt{N}} X_I, \;  P_I\rightarrow\frac{1}{\sqrt{N}} P_I,%
\end{equation}
For the $\mathrm{O}(2)$ model, we forgo the indexed notation and work with words built from the letters $\{Z, \bar Z, P, \bar P\}$.  
For practical implementation, it is convenient to define
$
\Pi \equiv -\, \i P,
$
so that all correlators appearing in the bootstrap are real~\cite{LinZheng1}.
\subsection{Constraints}
\def\O{\mathcal{O}}

The dynamical constraints include the equations of motion for a stationary state:
\begin{align}
    \langle [ H , \mathcal{O} ] \rangle = 0. \label{eom} \quad \forall \mathcal{O} \text{ single-trace operator}.
\end{align}

We refer to constraints that do not involve the explicit form of the Hamiltonian as ``kinematic constraints.'' These include cyclicity of the trace and gauge invariance. The ground states of these models are time-reversal symmetric, which, together with Hermiticity of the matrices, implies that $\ev{\tr \mathcal{O}_1 \cdots \mathcal{O}_n} = \pm \ev{\tr \mathcal{O}_n \cdots \mathcal{O}_1}$   where we choose $+(-)$ if there are an even (odd) number of $P$'s in the correlator~\footnote{Here $O_i$ are single letters, e.g., an $X_I$ or a $P_J$.}.

We also impose positivity of the inner product:
\begin{align}\label{posM}
    \mathcal{M}_{ij} = \ev{\tr \bar{\O}_i \O_j}, \quad \mathcal{M} \succeq 0 .
\end{align}
and ground state positivity:
\begin{equation}\label{gsp}
    \begin{split}
        \mathcal{N}_{ij} = \langle \tr  \bar{\O}_i [ H , \co_j ] \rangle, \quad \mathcal{N} \succeq 0.
    \end{split}
\end{equation}
This inequality immediately follows from demanding that the state $\co  \ket{\Omega}$ has an average energy larger than the ground state energy $E_0 = \bra{\Omega} H \ket{\Omega}$.
This can be viewed as the zero-temperature limit of the finite temperature bootstrap \cite{Araki:1977px, Fawzi:2023fpg, Cho:2024kxn, Cho:2025vws}.
Note that, here, we consider {\it adjoint} operators $\co_i, \co_j$ with two open matrix indices. This is justified since the ground state of the ungauged model is a gauge singlet. As in \cite{LinZheng1}, we decompose operators into irreps of $\mathrm{O}(D)$ and consider positivity for each irrep separately, as operators in different irreps are orthogonal due to group theoretical reason. For $D=2$, this is particularly convenient, since the irreps are labeled by charge; the letters $Z,P$ carry charge $1$ and $\bar{Z}, \bar{P}$ carry charge $-1$. For further details, we refer the reader to Appendices~\ref{app: example} and~\ref{app: irrep}, which contain an explicit implementation of the irreducible decomposition as well as a detailed worked example of all the constraints, and also include Refs. \cite{Cvitanovic:2008zz}.

\subsection{Hierarchy}
Following the approach developed in \cite{Kazakov:2024ool, LinZheng1}, we introduce a hierarchy among the set of all variables, and impose bootstrap constraints up to some level of the hierarchy.
The hierarchy is defined by sorting operators into {\it levels}: we assign the basic fields $\ell(X_I) = 1$, $\ell(P_I) = 2$. This assignment has the feature that $\ell([H,\mathcal{O}]) = \ell(\mathcal{O}) + 1$. 
We enumerate all variables up to a given level and take the quotient of this space by the equality constraints. The resulting ``search space'' can be parameterized by a choice of ``free variables'' which are the actual variables that enter the semidefinite program. Table~\ref{tab: freevariables} displays the number of free variables for each level. 

\begin{table}
\centering
\begin{tabular}{|c|cc|cc|}
\hline
\multirow{2}{*}{level}
  & \multicolumn{2}{c|}{$D=2$}
  & \multicolumn{2}{c|}{$D=9$} \\
\cline{2-5}
  & free variables & all variables
  & free variables & all variables \\ \hline
 4  &   3   & 14 &  3   & 7 \\
 6  &   8   & 94 &  10   & 52 \\
 8  &  22   & 614 &  43   & 487 \\
10  &  77   & 4086 & 289  & 5737 \\
12  & 326   & 27830 &  2859   & 81442 \\
14  & 1569  & 192374 &  ---   & 1348057 \\ \hline
\end{tabular}\caption{\label{tab: freevariables} The number of free variables in the semidefinite programming problem (after taking the quotient by the kinematic and dynamical constraints), for the $D=2$ and $D=9$ models.
}
\end{table}

\subsection{Nonlinear relaxation}
The cyclicity of the trace relates single-trace correlators to double-trace correlators.  
Since we wish to work at large $N$, we impose large-$N$ factorization.  
This implies that the single-trace correlation functions entering the matrices $\mathcal{M}$ and $\mathcal{N}$ are constrained by a set of quadratic relations.  
Consequently, the matrices $\mathcal{M}$ and $\mathcal{N}$ are, in general, nonlinear functions of the search-space parameters.

If the number of nonlinear parameters $\lesssim 3$, one can scan over them. %
However, at higher levels, the number of nonlinear parameters grows, and one should use the technique of nonlinear relaxation~\cite{Kazakov:2021lel, Kazakov:2022xuh}~\footnote{In principle, one can also consider other methods like the navigator function \cite{Reehorst:2021ykw}.}. 
The basic idea is to introduce a new variable for each double-trace that appears, e.g., $y =\ev{\tr P_I P_I }^2$. Then we may replace the equation $y = \ev{\tr P_J P_J}^2$ with the weaker (but rigorous) inequality $y \ge (\ev{\tr P_J P_J})^2$. This can be encoded in a positive semidefinite matrix:
$  \left(\begin{array}{cc}
1  & x \\
x & y \\
\end{array}\right) \succeq 0.$ In addition, we introduce some new positivity relations (beyond what was done in \cite{Kazakov:2021lel, Kazakov:2022xuh}) by imposing $y \le \ev{\tr (P_I P_I)(P_J P_J)}$ since positivity requires $\ev{\tr (P_I P_I)^2} \ge \ev{\tr P_I P_I }^2 $. 
We refer the reader to Appendix B for more detailed implementation. After performing this relaxation, we are left with a standard semidefinite programming problem with an expanded set of variables~\footnote{Although in principle we could relax all quadratic variables, we found it best to scan over $\ev{\tr X_I X_I}$ while relaxing the remaining quadratic variables.}.

\color{black}

\begin{table*}[ht]
  \centering
  \setlength{\tabcolsep}{8pt}
  \renewcommand{\arraystretch}{1.2}
  \small
  \begin{tabular}{@{} l l l l @{} }
    \toprule
    \textbf{Model } & $\mathcal{E}$ & $\langle \operatorname{tr} X_I X_I \rangle$ & level 4 operator \\
    \midrule
    \addlinespace[2pt]
   {\textbf{$M^2=0,\;D=2$}} & & &  $\ev{\tr (Z^2 \bar{Z}^2)}$  \\
Bootstrap   & {\color{blue} $[0.707832,\,0.707868]$} & %
    {\color{blue} [1.15420, 1.15460]}
    & %
    {\color{blue} [0.37055, 0.37085]}   \\
    Monte Carlo \cite{Bodendorfer:2024egw} & $0.7039(11)$ & --- & --- \\
     Loop truncation \cite{Mathaba:2023non} & $0.7056(2)$ & 1.172(1) & 0.383(2) \\
     Large $D$ expansion \cite{Mandal:2009vz} & $0.756$ & $0.985$ & 0.177 \\
    \hline
    \addlinespace[4pt]
{\textbf{$M^2=1,\;D=2$}} & & & $\ev{\tr (Z^2 \bar{Z}^2)}$  \\
    Bootstrap   & {\color{blue} [1.172098376, 1.172098408]}%
    &%
    {\color{blue} [0.77800898, 0.77800934]}%
    
    & {\color{blue} [0.15850588, 0.15850607]} \\
    
    Monte Carlo \cite{Bodendorfer:2024egw} & $1.1654(11)$ & ---   & --- \\
    Loop truncation \cite{Mathaba:2023non} & $1.17198$ & $0.7784$ &  0.1588 \\
    \hline
    \addlinespace[4pt]
{\textbf{$M^2=0,\;D=9$}} & & & $\ev{\tr X_I X_I X_J X_J }$  \\
    Bootstrap   & $[6.69946,6.69968]$ &%
    
    $[2.29195,2.29218]$ & %
    
    $[5.7787,5.7804]$ \\
    
    Monte Carlo \cite{Kawahara:2007fn}  & $6.695(5)$ & $2.291(1)$ & --- \\
    Large $D$ expansion \cite{Mandal:2009vz} & 6.713 & 2.279 & 5.646 \\
    \bottomrule
  \end{tabular}
  \caption{Ground state energy $\mathcal{E}$ bounds and expectation values of $\langle \operatorname{tr} X_I X_I \rangle$ from bootstrap and Monte Carlo studies for various masses $M^2$ and dimensions $D$. Results in {\color{blue} blue} are level 14 bootstrap bounds; the $D=9$ bootstrap results are level 11. Our rigorous results are complementary to the finite temperature 
  Monte Carlo results as they do not involve any continuum or large $N$ extrapolation.
  }
  \label{tab:results}
\end{table*}

\begin{figure*}[ht]
\centering
\includegraphics[width=0.50\textwidth]{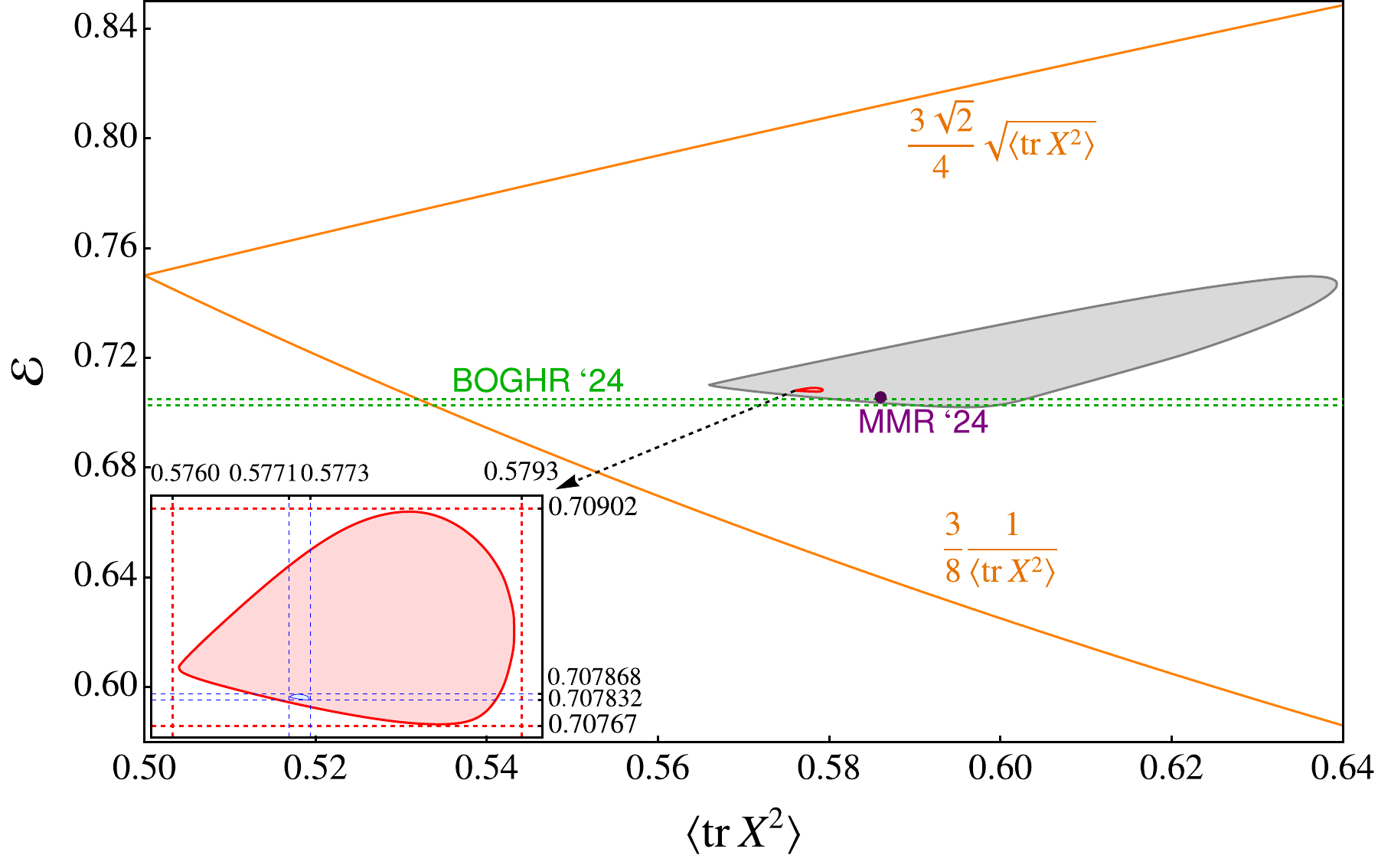}
\includegraphics[width=0.48\textwidth]{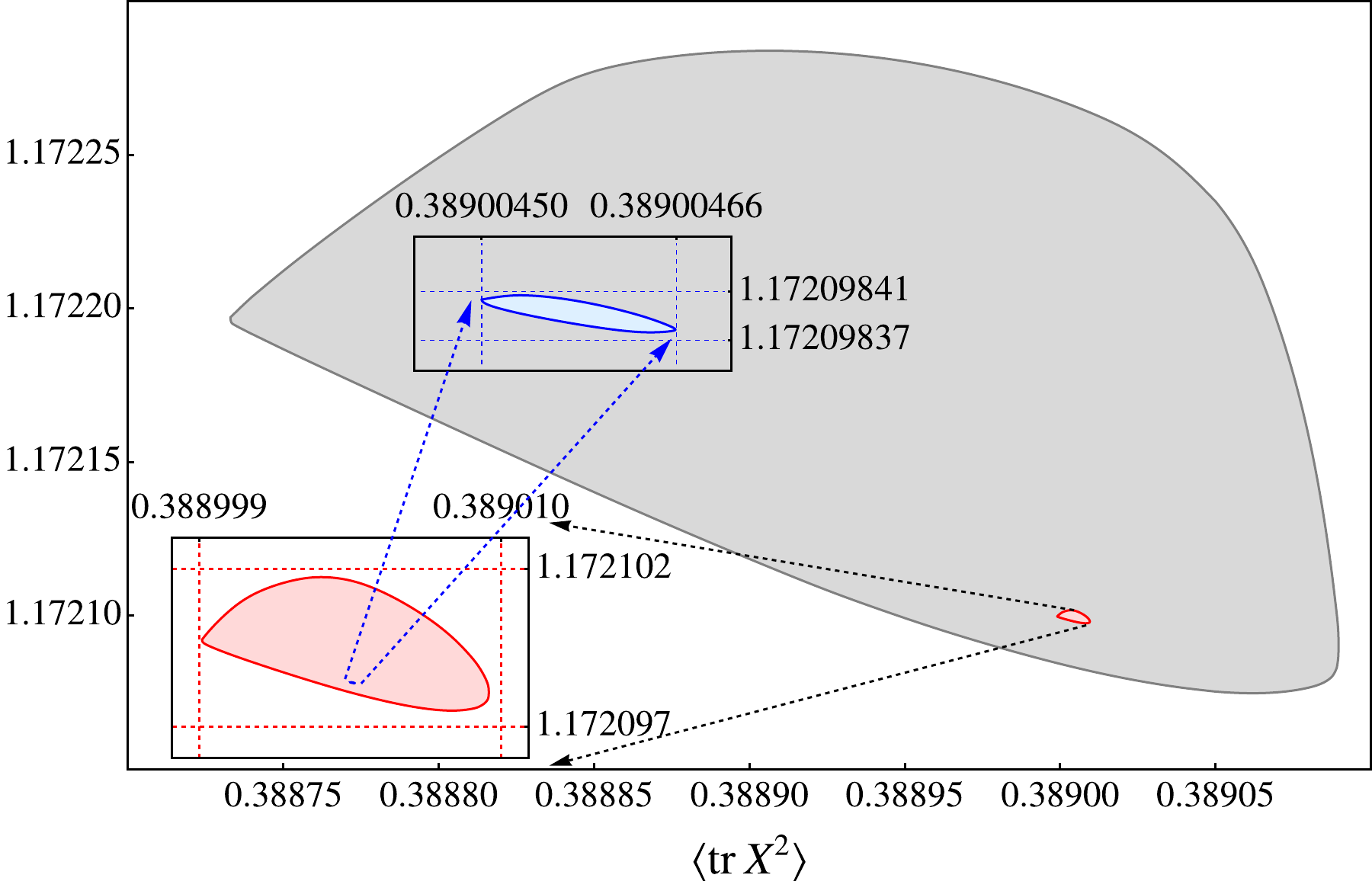}
\caption{\label{fig: massless} 
{\it Left}: Bootstrap constraints on the ground state/confining phase energy $\mathcal{E} = E_0 /(g^{2}_\ym N)^{1/3} $ and the simplest observable $\ev{\tr X^2} = \hf  \ev{\tr X_I X_I} =  \ev{\tr Z \bar{Z}}$ for the $D=2$ massless matrix model. 
The {\color{orange} level 6} constraint \eqref{eq: analytic} for the massless $D=2$ model has the shape of a peninsula. The feasible island first appears at {\color{gray} level 10}. The island shrinks rapidly as we go to higher levels, so we show it more clearly in the inset panel. The small {\color{blue} level 14} island is not visible to the naked eye in the main figure. Comparison to previous numerical results in the literature \cite{Bodendorfer:2024egw} ({\color{dgreen} green}) and \cite{Mathaba:2023non} ({\color{purple} purple}) are given. 
{\it Right}: the {\color{gray} level 10}, {\color{red} 12}, and $\color{blue} 14$ constraints for the massive $D=2$ model with $M^2=1$. The {\color{gray} level 10 island} already excludes the Monte Carlo result \cite{Bodendorfer:2024egw}. A potential future application of these tiny islands is to enable a better estimate of potential systematic uncertainties in Monte Carlo simulations. }
\end{figure*}

\section{Bootstrap results \label{sec: results}}

\begin{figure}
\includegraphics[width=0.45\textwidth]{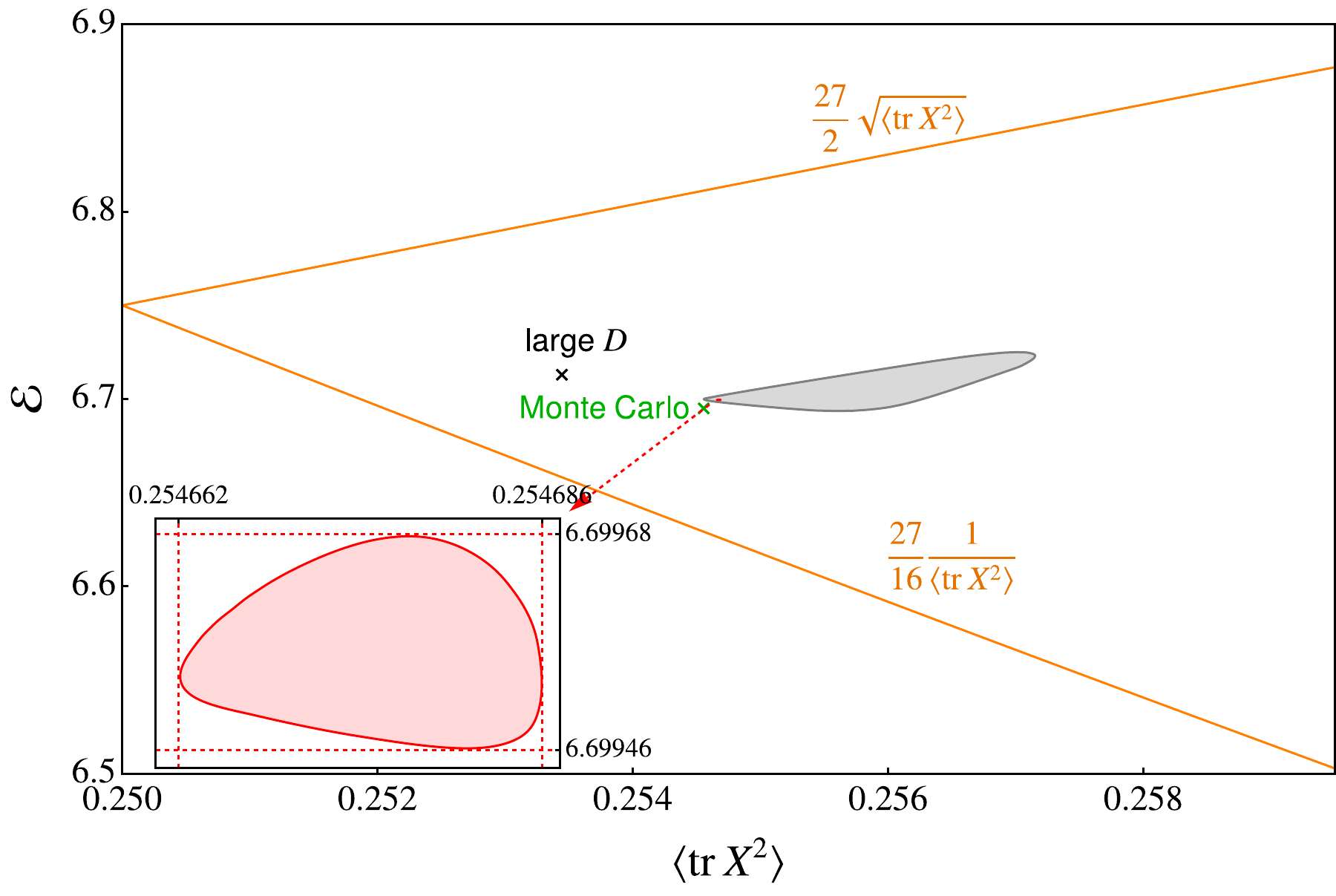}
\caption{\label{fig: masslessD9} 
Here we display the level 10 constraints for the massless $D=9$ model, sometimes referred to as ``bosonic BFSS''. The cross $\times$ indicates the estimate from the $1/D$ expansion \cite{Mandal:2009vz} and the {\color{dgreen} green $\times$} is from the $N=32$ Monte Carlo simulation in \cite{Kawahara:2007fn}, see also \cite{Filev:2015hia}. The orange curves are the analytic bounds \eqref{eq: analytic}. The level 10 island is displayed in {\color{gray} gray}; the  red island (barely visible) near the western tip is the {\color{red} level 11} island; it is displayed more clearly in the inset panel.} 
\end{figure}

We considered the bootstrap for both the massless model ($M=0$) and the massive model with $M^2 = 1$ (equivalently, $\lambda_\text{eff} = 1$).  For the case $D=2$, we performed a bootstrap analysis up to level~14, while for $D=9$ the analysis was carried out up to level~11. We mainly consider the bounds for the dimensionless normalized ground state energy $\mathcal{E} \equiv \frac{E_0}{N^2} (g^2_\text{YM} N)^{-1/3}$ and $\ev{\tr X^2} = \frac{1}{D}\ev{\tr X_I X_I}$ at different levels. The results are presented in Table~\ref{tab:results}, Figs.~\ref{fig: massless} and~\ref{fig: masslessD9}.

For the $D=2$ model, \cite{Rinaldi:2021jbg, Bodendorfer:2024egw} performed a Euclidean lattice computation using the Hybrid Monte Carlo algorithm. For $D=2$, a large $N$ and continuum extrapolation yielded the numbers that we display in Table \ref{tab:results}. For $D=9$, one may compare to the $N=32$ Monte Carlo lattice simulation of \cite{Kawahara:2007fn}.  We also compare\footnote{Our $\lambda_\text{eff} = 1$ becomes $m=2, g=2$ in the conventions of \cite{Mathaba:2023non}, after a canonical transformation.} to the interesting numerical approach of \cite{Mathaba:2023non}. Their approach is based on a truncation of the loop equations \cite{Jevicki:1982jj, Jevicki:1983wu, Rodrigues:1985aq, Koch:2021yeb, Mathaba:2023non}.

We also bootstrapped the level 4 operator $\ev{\tr Z^2 \bar{Z}^2}  = \qrt \sum_{I,J} \ev{\tr X_I X_J X_I X_J} $. Together with the constraints, the variables in Table~\ref{tab:results} fully determine the expectation value of all operators up to level 4. 
At level 8, this operator is bounded from both sides for the massive $M^2 = 1$ model (and is unbounded from above at level 6 or below). To find a two-sided bound for the massless model, we had to go to level 10.

To build intuition for these bounds, we note that at lower levels, it is possible to analyze the bootstrap constraints by hand and identify which inequalities are saturated. %
At level $5$, we may derive a simple analytic bound on $\mathcal{E}$ and $\ev{\tr X^2}$\footnote{see \cite{Lin:2023owt} and \cite{LinZheng1} for similar bounds in the BFSS case.}. From the concrete example presented in Appendix~\ref{app: example}, we obtain the following bounds:
\begin{small}
\begin{equation}\label{eq: analytic}
\begin{split}
    \frac{\mathcal{E}}{D} &\ge \max\Big\{ \frac{3  }{16 \ev{\tr X^2} }+\frac{ M^2   \ev{\tr X^2}}{4},M^2 \ev{\tr X^2}\Big\},\\
    \frac{\mathcal{E}}{D} &\le 
  \frac{1}{8} \left[ 2 M^2  \ev{\tr X^2} + 3   \left(2 (D-1) \ev{\tr X^2}+M^2 \right)^{1/2}  \right]
\end{split}
\end{equation}
\end{small}

For the massless case,  this yields $\frac{1}{(8 D-8)^{1/3} }  \le \ev{\tr X^2}$.
Surprisingly, for $M^2= 0$, adding the bootstrap constraints up to level 8 did not lead to any improvement for these inequalities. In other words, \eqref{eq: analytic} is the optimal bound up to level 8. The peninsula bound derived here is quite analogous to the level 5 peninsula in the BFSS case. The behavior of these bounds is markedly different for the massive case  $M^2 >0$. For massive models, the inequalities in \eqref{eq: analytic} carve out a compact allowed region (an “island”) already at level 5; the island shrinks at higher levels.

\section{Discussion \label{sec: discuss}}

At level $L < 10$, our results for the massless $M^2 = 0$ cases are qualitatively similar to the plots that were obtained in \cite{LinZheng1} for the maximally supersymmetric Yang-Mills matrix model (the D0 brane or BFSS matrix model). 
To make the comparison more concrete, note that in the SUSY case, there are three terms in the Hamiltonian $E = K +V +F$ whereas in the bosonic models there are only two terms $E = K+V$. These terms are related by the virial theorem, $-2K + 4V +F =0$ in the SUSY case and $-2K + 4V = 0$ in the bosonic (massless) case.
However, SUSY gives an additional constraint $E = 0$ for the ground state of BFSS, so there is really only one independent variable ``energy''-like variable $\mathcal{E}$ in both the SUSY and non-SUSY cases.
In addition to this level 4 variable (which can be viewed as the kinetic or potential energy), there is the level 2 variable $\ev{\tr X^2}$ and one other level 4 variable $\ev{\tr X_I X_I X_J X_J}$.
In all models that we studied, a peninsula in the $\ev{\tr X^2}-\mathcal{E}$ appears exactly at level 5: there is an unconditional lower bound on $\ev{\tr X^2}$ at level 5, but no upper bound. Furthermore, up to level 10, there is a ``vertical'' peninsula in the $\ev{\tr X^2} - \ev{\tr X_I X_I X_J X_J}$ plot that resembles figure 3 in \cite{LinZheng1}.

These features are special to the $M^2 = 0$ models. When we added a mass term, we saw that the bootstrap behaved quite differently; an island appears already at level 4. It seems that the ``classically'' flat directions present in these potentials make it more difficult to upper bound the size of the ground state wavefunction $\ev{\tr X^2}$.

At level 10, the feasible region becomes an island in the massless $D=2$ and $D=9$ cases we studied in this work. We are therefore optimistic that an island will appear at level 10 (or perhaps within a few levels beyond) in the supersymmetric model. Note that in the supersymmetric setting, ground state positivity \eqref{gsp} was implicitly imposed by the supercharge equations of motion \cite{LinZheng2}.

An extension of our work to the future is to study the models at finite temperature, using the approach outlined in \cite{Cho:2024kxn}. (One could also study observables in a microcanonical window of energies $E/N^2$ above the ground state.) In fact, the near agreement between our zero temperature result and the low (but $O(1)$ in 't Hooft units) temperature Monte Carlo results may already be viewed as strong evidence that these theories are in a confined phase, where simple observables are expected to be independent of temperature at leading order in $1/N^2$. 
Another remote but interesting possibility is that the small discrepancy between our zero temperature results and the finite temperature Monte Carlo results is {\it not} due to underestimated systematic error but, instead, represents a failure of the conjectured phase diagram. In other words, to compare our zero temperature bootstrap result with the low-temperature Monte Carlo result, we assumed that the theory is in a confined phase where to leading order in $1/N^2$ the thermal energy is independent of temperature. This assumption is supported by the Monte Carlo simulations and Eguchi-Kawai equivalence \cite{Kawahara:2007fn, Bodendorfer:2024egw, Rinaldi:2021jbg} but remains unproven analytically.
The existence of a confined phase is expected for large $D$ from the $1/D$ expansion \cite{Mandal:2009vz}, but, here, we can test the conjectured phase diagram even at $D=2$.

 Another interesting problem is to compute the Polyakov loop using the bootstrap formalism. This could be done by modifying the gauge constraint.  Combining these two future directions, one could confirm the existence of the Gross-Wadia-Witten transition at finite $D$ \cite{Aharony:2004ig, Kawahara:2007fn, Mandal:2009vz}. One could also study time-dependent correlators \cite{Cho:2025vws} with the eventual goal of computing the quasinormal modes \cite{Dodelson:2024atp, Dodelson:2025rng}.

Finally, it would be interesting to study supersymmetric models with SO($D$) symmetry, see \cite{Kim:2006wg, Rinaldi:2021jbg}. These models have fewer supersymmetries than the BFSS model; their gravity duals are much less understood compared to BFSS \cite{Banks:1996vh, Itzhaki:1998dd}. In particular, the $\mathcal{N}=2$ SUSY model \cite{Hoppe:1997wba, Hoppe:1997fr} would be an interesting bootstrap target. It would be interesting to test whether there are metastable bound states in the planar 't Hooft limit of these models, which would be akin to the D0 brane black hole in the BFSS model.
\begin{acknowledgments}
\section*{Acknowledgments}

We thank Yiming Chen, Minjae Cho, Mathew Dodelson, Barak Gabai, Davide Gaiotto, Masanori Hanada, Igor Klebanov, Stephen Shenker, Jessica Yeh, and Xi Yin for discussions. We especially thank Davide Gaiotto for providing computational resources that enabled most of the results presented in this work. H.L. is supported by a Bloch Fellowship and by NSF Grant PHY-2310429. Z.Z. is supported by the Simons Foundation grant \#994308 for the Simons Collaboration on Confinement and
QCD Strings.
\end{acknowledgments}

\clearpage
\onecolumngrid
\appendix
\newcounter{appsection}
\renewcommand{\theappsection}{\Alph{appsection}}
\providecommand{\theHappsection}{}
\renewcommand{\theHappsection}{appendix.\arabic{appsection}}
\setcounter{secnumdepth}{2}
\renewcommand{\thesubsection}{\Alph{appsection}.\arabic{subsection}}
\renewcommand{\theHsubsection}{appendix.\Alph{appsection}.\arabic{subsection}}
\renewcommand{\theequation}{\Alph{appsection}\arabic{equation}}
\renewcommand{\theHequation}{appendix.\Alph{appsection}.\arabic{equation}}
\newcommand{\appsection}[2]{%
  \refstepcounter{appsection}\label{#2}%
  \setcounter{equation}{0}%
  \setcounter{subsection}{0}%
  \section*{Appendix \Alph{appsection}: #1}%
}

\appsection{Numerical implementation}{app: details}

To solve the semidefinite programming (SDP) constructed, we used \texttt{SDPA-GMP}, an arbitrary-precision arithmetic SDP solver~\cite{SDPA-GMP}. We observed that our SDP problem exhibited significant numerical instability, which motivated our choice of an arbitrary-precision solver. These numerical instabilities can be traced back to the lack of strict feasibility in the primal form of the SDP. For further technical discussion of this type of numerical instability, and possible (though not systematic) strategies for mitigating it, we refer the reader to, for example,~\cite{Permenter_2017}.

Our \href{https://github.com/Canonical111/O2massiveBootstrap}{Github repository} includes a \textsc{Mathematica} notebook as an example for generating the final SDP. Users are free to export the output to other SDP solvers to obtain bounds on specific variables. The built-in solver in \textsc{Mathematica} may suffice for lower bootstrap levels, but for more ambitious computations, a higher-precision solver is necessary.

\appsection{Relaxation}{app: relax}
In this appendix, we provide additional details on the implementation of the relaxation procedure.  
Let us denote by $\{\mathcal{O}_i\}$ the set of singlet operators that appear up to a given truncation level.  
(Here, each $\mathcal{O}_i$ must be a singlet under the global symmetry; otherwise its expectation value vanishes.)  
We then define the corresponding column vector of expectation values,
\begin{equation}
    x_i \equiv \left\langle \operatorname{tr} \mathcal{O}_i \right\rangle,
\end{equation}
so that $\vec{x}$ collects the expectation values of all singlet operators included at this level. The quadratic variables are then given by the elements of the following matrix $Q$, with entries:
\begin{equation}\label{eq:nonconvex}
    Q_{ij} = x_i x_j,
\end{equation}
or, more compactly,
\begin{equation}
    Q = \vec{x}\vec{x}^{\mathrm{T}}.
\end{equation}
Whenever quadratic terms appear in our constraints, we replace them with the corresponding elements of the matrix $Q$. In this way, we isolate the nonconvexity of our optimization problem in~\eqref{eq:nonconvex}. In the following, we study the relaxation of~\eqref{eq:nonconvex}. 

We observe that the matrix $Q$ satisfies the following two convex conditions. First, the matrix
\begin{equation}
    \begin{pmatrix}
    1 & \vec{x}^{\mathrm{T}} \\
    \vec{x} & Q
    \end{pmatrix} \succeq 0
\end{equation}
is positive semidefinite, since it is the outer product of the vector $(1,\, \vec{x}^{\mathrm{T}})^{\mathrm{T}}$ with itself.

Additionally, the following matrix is also positive semidefinite, as implied by~\eqref{posM}:
\begin{equation}
    \left\langle \operatorname{tr} \left( \mathcal{O}_i - \langle \mathcal{O}_i \rangle \right)^\dagger 
    \left( \mathcal{O}_j - \langle \mathcal{O}_j \rangle \right) \right\rangle 
    = \mathcal{M}_{ij} - Q_{ij} \succeq 0.
\end{equation}

In the relaxation, we introduce a new matrix $\mathcal{Q}$, which, instead of satisfying~\eqref{eq:nonconvex}, is required to satisfy the following two conditions:
\begin{equation}\label{eq: relax1}
    \begin{pmatrix}
    1 & \vec{x}^{\mathrm{T}} \\
    \vec{x} & \mathcal{Q}
    \end{pmatrix} \succeq 0,
\end{equation}
\begin{equation}\label{eq: relax2}
    \mathcal{M} \succeq \mathcal{Q}.
\end{equation}
We then replace every occurrence of an element of $Q$ in other constraints with the corresponding element of $\mathcal{Q}$.

The relaxation implementation described here, namely \eqref{eq: relax1} and \eqref{eq: relax2}, is stronger than the previous approach in~\cite{Kazakov:2021lel, Kazakov:2022xuh}, where only \eqref{eq: relax1} was imposed.

\appsection{Large \texorpdfstring{$D$}{D} comparison}{app: largeD-main}
One may also compare our results in the main text with the large-$D$ approximation of this model; see~\cite{Hotta:1998en, Mandal:2009vz, Anegawa:2024qoz}. 
In this limit, the system can be viewed as a deformation of an O($D$) vector model, allowing for a controlled resummation of bubble diagrams. Details of the large-$D$ resummation are provided in Appendix~\ref{app: largeD}.

In our conventions, this gives the estimates

\begin{small}
\begin{equation}\label{largeD}
    \begin{split}
     &   \ev{\tr X_I X_I}  = \hf D^{2/3} \left[ 1 + \frac{2}{D} \left(\frac{7 \sqrt{5}}{30} - \frac{9}{32}\right)  + \cdots \right]  \approx 2.279  \\
    &\mathcal{E}  = D^{4/3} \left[\frac38 + \frac{1}{D}\left( \frac{\sqrt{5}}{2}-\frac{81}{64} \right) + \cdots  \right] \approx 6.713  \\
  &\ev{\tr X_I X_I X_J X_J} =   D^{4/3} \left [\qrt  + \frac{1}{D} \left( \frac{\sqrt{5}}{3}-\frac{9}{32} \right)  + \cdots\right]  \approx 5.646 
\end{split}
\end{equation}
\end{small}

The first two were computed in \cite{Mandal:2009vz}; the third, to our knowledge, has not appeared in the literature and is derived in Appendix~\ref{app: largeD}. According to the Virial theorem, the potential energy $2V = \ev{\tr X_I^2 X_J^2}-\ev{\tr X_I X_J X_I X_J} = 2\mathcal{E}/3 $. We observe that to leading order in $D$, $\ev{\tr X_I^2 X_J^2} = 2 \mathcal{E}/3$. This implies that the ``alternating correlator'' $\ev{\tr X_I X_J X_I X_J}$ at leading order in $N$ is sub-leading in $1/D$ compared to $\ev{\tr X_I^2  X_J^2}$. 

We have reported the numbers for $D=9$ in \eqref{largeD}. Although it is a bit extreme, one can also try $D=2$; one obtains estimates which are unsurprisingly less accurate than for $D=9$. It would be interesting to repeat the bootstrap at various values of $D$ to better test the large $D$ scaling. Some results for $D=3$ using the loop truncation method have already been obtained \cite{Rodrigues:2025sbu}.

\appsection{The adjoint sector}{app: adjoint}
Let us comment on gauge non-singlet states in matrix quantum mechanics. A basic quantity is the energy gap to the adjoint sector $\Delta E_\text{adj}$. This quantity is related to the adjoint Polyakov loop at low temperatures $P \sim   e^{-\beta \Delta E_\text{adj}}$. Here we present a simple bootstrap bound:
\begin{small}
    \begin{equation}\label{eq:time_dependent_bd}
        \begin{split}
            \Delta \mathcal{E}_\text{adj} &=   \Delta {E}_\text{adj} (g^2_\ym N)^{-1/3} \leq \frac{\ev{\tr X_I[H,X_I]}}{\ev{\tr X_I X_I}} = \frac{1}{2 \ev{\tr X^2}}
        \end{split}
    \end{equation}
\end{small}
where we have used the ground state positivity matrix \eqref{gsp}. 
(We also give a different derivation of this bound in \cite{Cho:2025vws}, by bootstrapping time-dependent correlators.)

Combined with our level 11 lower bound on $\ev{\tr X^2}$, we have the following comparison with the Monte Carlo estimate~\cite{Berkowitz:2018qhn}:
\begin{align} \label{adjNum}
        \Delta \mathcal{E}_\text{adj} \le  1.96339 , \quad \Delta \mathcal{E}_\text{adj}^\text{Monte Carlo} \approx 2.043(76).
\end{align}
This estimate applies to the $D=9, M=0$ model.
The Monte Carlo estimate~\cite{Berkowitz:2018qhn} was obtained by contrasting the gauged and ungauged thermodynamics; \cite{Berkowitz:2018qhn} also reports an alternative estimate based on measuring $\langle \operatorname{tr} X^2 \rangle$ in the ungauged model, yielding
\[
\Delta \mathcal{E}_\text{adj}^{\text{Monte Carlo}} \lesssim 1.936(71).
\]
At first sight, it may seem surprising that the relatively simple bound~\eqref{eq:time_dependent_bd} essentially reproduces the Monte Carlo value.  The reason is that the bound~\eqref{eq:time_dependent_bd} becomes saturated in the large-$D$ limit.  At infinite $D$, the state $X_{ij} \ket{\Omega}$ is an energy eigenstate; this follows from the expressions for the propagator in Appendix~\ref{app: largeD}, which take the form of a massive harmonic oscillator.

In the supersymmetric BFSS model~\cite{Maldacena:2018vsr, Berkowitz:2018qhn}, our rigorous lower bounds imply
\[
\Delta \mathcal{E}_\text{adj} \le 1.41\, (g_\text{YM}^2 N)^{1/3}.
\]
If we incorporate the Monte Carlo result for $\langle \operatorname{tr} X^2 \rangle \approx 0.378$~\cite{Pateloudis:2022ijr}, which lies slightly above our level-9 lower bound~\cite{LinZheng1}, the estimate improves to
\[
\Delta \mathcal{E}_\text{adj} \lesssim 1.3.
\]
Meanwhile, the dedicated Monte Carlo study~\cite{Berkowitz:2018qhn} finds
\[
\Delta \mathcal{E}_\text{adj} \approx 0.92(11)\, (g_\text{YM}^2 N)^{1/3}.
\]
Taken together, the Monte Carlo results~\cite{Pateloudis:2022ijr, Berkowitz:2018qhn} indicate that—unlike the bosonic $D=9$ model—the supersymmetric theory does not come close to saturating the general bound~\eqref{eq:time_dependent_bd}.  This reflects the genuinely nontrivial dynamics governing the low-energy limit of the supersymmetric BFSS model.

In fact, the simple bound \eqref{eq:time_dependent_bd} can be improved by simply considering operators other than $X_I$. For example, one can consider both $X_I$ and $P_I$ and use the full ground state positivity matrix \eqref{gsp}; this leads to an improved bound:
\begin{equation}
    \Delta\mathcal{E}_\text{adj} \le -\frac{1}{4 p x-1}\left(\sqrt{\left(p-2 (D-1) x^2\right)^2 \\-(4 p x-1) \left(2 (D-1) x-4 p^2\right)}-2 (D-1) x^2+p\right).
\end{equation}
with $p= \ev{\tr P_I P_I}/D, x = \ev{\tr X_I X_I}/D$.
 A method to systematically bound the gap will be reported in upcoming work \cite{ZhengInPrep}, we hope to apply this method to BFSS in the future \cite{LinZheng2}.

\appsection{A worked example: level 5}{app: example}

In this section, we present a detailed worked example of the Hamiltonian discussed in the main text:
\begin{align}
    H &= \frac{1}{2} \sum_{I=1}^D \left( -\Tr \Pi_I \Pi_I + M^2  \Tr X_I X_I \right) 
        - \frac{1}{4} \sum_{I,J = 1} ^D \Tr \left[ X_I , X_J \right]^2 \\
    C &= \sum_{I=1}^D \left( [X_I, \Pi_I] - \mathbf{1} \right)
\end{align}
Here, we have introduced $\Pi_I = -\i P_I$. The advantage of this replacement is that the correlators of words composed of the letters $\Pi_I$ and $X_I$ are real numbers.

\subsection{Equations}
Up to level 5, we have the following $19$ variables:
\begin{align}
    \begin{split}
        &\left\langle\mathop{\tr} \Pi_{\text{I}}\Pi_{\text{I}}\right\rangle ,\left\langle\mathop{\tr} \Pi_{\text{I}}X_{\text{I}}\right\rangle ,\left\langle\mathop{\tr} X_{\text{I}}\Pi_{\text{I}}\right\rangle ,\left\langle\mathop{\tr} X_{\text{I}}X_{\text{I}}\right\rangle ,\left\langle\mathop{\tr} \Pi_{\text{I}}X_{\text{I}}X_{\text{J}}X_{\text{J}}\right\rangle ,\left\langle\mathop{\tr} \Pi_{\text{I}}X_{\text{J}}X_{\text{I}}X_{\text{J}}\right\rangle ,\\
        &\left\langle\mathop{\tr} \Pi_{\text{I}}X_{\text{J}}X_{\text{J}}X_{\text{I}}\right\rangle ,\left\langle\mathop{\tr} X_{\text{I}}\Pi_{\text{I}}X_{\text{J}}X_{\text{J}}\right\rangle ,\left\langle\mathop{\tr} X_{\text{I}}\Pi_{\text{J}}X_{\text{I}}X_{\text{J}}\right\rangle ,\left\langle\mathop{\tr} X_{\text{I}}\Pi_{\text{J}}X_{\text{J}}X_{\text{I}}\right\rangle ,\left\langle\mathop{\tr} X_{\text{I}}X_{\text{I}}\Pi_{\text{J}}X_{\text{J}}\right\rangle ,\\
        &\left\langle\mathop{\tr} X_{\text{I}}X_{\text{I}}X_{\text{J}}\Pi_{\text{J}}\right\rangle ,\left\langle\mathop{\tr} X_{\text{I}}X_{\text{I}}X_{\text{J}}X_{\text{J}}\right\rangle ,\left\langle\mathop{\tr} X_{\text{I}}X_{\text{J}}\Pi_{\text{I}}X_{\text{J}}\right\rangle ,\left\langle\mathop{\tr} X_{\text{I}}X_{\text{J}}\Pi_{\text{J}}X_{\text{I}}\right\rangle ,\left\langle\mathop{\tr} X_{\text{I}}X_{\text{J}}X_{\text{I}}\Pi_{\text{J}}\right\rangle ,\\
        &\left\langle\mathop{\tr} X_{\text{I}}X_{\text{J}}X_{\text{I}}X_{\text{J}}\right\rangle ,\left\langle\mathop{\tr} X_{\text{I}}X_{\text{J}}X_{\text{J}}\Pi_{\text{I}}\right\rangle ,\left\langle\mathop{\tr} X_{\text{I}}X_{\text{J}}X_{\text{J}}X_{\text{I}}\right\rangle
    \end{split}
\end{align}
The notation for the variables used here follows that introduced in~\cite{LinZheng1}, where repeated indices $I, J, K, \ldots$ are implicitly summed over.

As discussed in the main text, these variables are subject to both kinematic and dynamical constraints. Below, we list the linear combinations of variables that vanish as a consequence of these constraints:
\begin{small}
\begin{equation}
D+\left\langle\mathop{\tr} \Pi_{\text{I}}X_{\text{I}}\right\rangle
-\left\langle\mathop{\tr} X_{\text{I}}\Pi_{\text{I}}\right\rangle = 0
\end{equation}
\begin{equation}
-\left\langle\mathop{\tr} \Pi_{\text{I}}X_{\text{I}}\right\rangle
+\left\langle\mathop{\tr} X_{\text{I}}\Pi_{\text{I}}\right\rangle -D = 0
\end{equation}
\begin{equation}
\left\langle\mathop{\tr} \Pi_{\text{I}}X_{\text{I}}\right\rangle
+\left\langle\mathop{\tr} X_{\text{I}}\Pi_{\text{I}}\right\rangle = 0
\end{equation}
\begin{equation}
-\left\langle\mathop{\tr} \Pi_{\text{I}}X_{\text{J}}X_{\text{J}}X_{\text{I}}\right\rangle
+\left\langle\mathop{\tr} X_{\text{I}}\Pi_{\text{I}}X_{\text{J}}X_{\text{J}}\right\rangle
-D\left\langle\mathop{\tr} X_{\text{I}}X_{\text{I}}\right\rangle = 0
\end{equation}
\begin{equation}
-\left\langle\mathop{\tr} X_{\text{I}}X_{\text{I}}\right\rangle
-\left\langle\mathop{\tr} \Pi_{\text{I}}X_{\text{J}}X_{\text{I}}X_{\text{J}}\right\rangle
+\left\langle\mathop{\tr} X_{\text{I}}\Pi_{\text{J}}X_{\text{I}}X_{\text{J}}\right\rangle = 0
\end{equation}
\begin{equation}
-\left\langle\mathop{\tr} X_{\text{I}}X_{\text{I}}\right\rangle
-\left\langle\mathop{\tr} \Pi_{\text{I}}X_{\text{I}}X_{\text{J}}X_{\text{J}}\right\rangle
+\left\langle\mathop{\tr} X_{\text{I}}\Pi_{\text{J}}X_{\text{J}}X_{\text{I}}\right\rangle = 0
\end{equation}
\begin{equation}
\left\langle\mathop{\tr} X_{\text{I}}\Pi_{\text{I}}X_{\text{J}}X_{\text{J}}\right\rangle
+\left\langle\mathop{\tr} X_{\text{I}}X_{\text{I}}\Pi_{\text{J}}X_{\text{J}}\right\rangle = 0
\end{equation}
\begin{equation}
\left\langle\mathop{\tr} X_{\text{I}}X_{\text{I}}\Pi_{\text{J}}X_{\text{J}}\right\rangle
-\left\langle\mathop{\tr} X_{\text{I}}\Pi_{\text{J}}X_{\text{J}}X_{\text{I}}\right\rangle = 0
\end{equation}
\begin{equation}
\left\langle\mathop{\tr} X_{\text{I}}X_{\text{I}}\right\rangle
+\left\langle\mathop{\tr} \Pi_{\text{I}}X_{\text{J}}X_{\text{J}}X_{\text{I}}\right\rangle
-\left\langle\mathop{\tr} X_{\text{I}}X_{\text{I}}X_{\text{J}}\Pi_{\text{J}}\right\rangle
+D\left\langle\mathop{\tr} X_{\text{I}}X_{\text{I}}\right\rangle = 0
\end{equation}
\begin{equation}
\left\langle\mathop{\tr} \Pi_{\text{I}}X_{\text{I}}X_{\text{J}}X_{\text{J}}\right\rangle
+\left\langle\mathop{\tr} X_{\text{I}}X_{\text{I}}X_{\text{J}}\Pi_{\text{J}}\right\rangle = 0
\end{equation}
\begin{equation}
-\left\langle\mathop{\tr} X_{\text{I}}X_{\text{I}}\Pi_{\text{J}}X_{\text{J}}\right\rangle
+\left\langle\mathop{\tr} X_{\text{I}}X_{\text{I}}X_{\text{J}}\Pi_{\text{J}}\right\rangle
-D\left\langle\mathop{\tr} X_{\text{I}}X_{\text{I}}\right\rangle = 0
\end{equation}
\begin{equation}
\left\langle\mathop{\tr} \Pi_{\text{I}}X_{\text{I}}X_{\text{J}}X_{\text{J}}\right\rangle
+\left\langle\mathop{\tr} X_{\text{I}}\Pi_{\text{I}}X_{\text{J}}X_{\text{J}}\right\rangle
+\left\langle\mathop{\tr} X_{\text{I}}X_{\text{I}}\Pi_{\text{J}}X_{\text{J}}\right\rangle
+\left\langle\mathop{\tr} X_{\text{I}}X_{\text{I}}X_{\text{J}}\Pi_{\text{J}}\right\rangle = 0
\end{equation}
\begin{equation}
\left\langle\mathop{\tr} X_{\text{I}}X_{\text{J}}\Pi_{\text{I}}X_{\text{J}}\right\rangle
-\left\langle\mathop{\tr} X_{\text{I}}\Pi_{\text{J}}X_{\text{I}}X_{\text{J}}\right\rangle = 0
\end{equation}
\begin{equation}
\left\langle\mathop{\tr} X_{\text{I}}\Pi_{\text{J}}X_{\text{I}}X_{\text{J}}\right\rangle
+\left\langle\mathop{\tr} X_{\text{I}}X_{\text{J}}\Pi_{\text{I}}X_{\text{J}}\right\rangle = 0
\end{equation}
\begin{equation}
-\left\langle\mathop{\tr} X_{\text{I}}X_{\text{I}}\right\rangle
+\left\langle\mathop{\tr} X_{\text{I}}X_{\text{I}}X_{\text{J}}\Pi_{\text{J}}\right\rangle
-\left\langle\mathop{\tr} X_{\text{I}}X_{\text{J}}\Pi_{\text{J}}X_{\text{I}}\right\rangle = 0
\end{equation}
\begin{equation}
\left\langle\mathop{\tr} X_{\text{I}}X_{\text{J}}\Pi_{\text{J}}X_{\text{I}}\right\rangle
-\left\langle\mathop{\tr} X_{\text{I}}\Pi_{\text{I}}X_{\text{J}}X_{\text{J}}\right\rangle = 0
\end{equation}
\begin{equation}
\left\langle\mathop{\tr} X_{\text{I}}\Pi_{\text{J}}X_{\text{J}}X_{\text{I}}\right\rangle
+\left\langle\mathop{\tr} X_{\text{I}}X_{\text{J}}\Pi_{\text{J}}X_{\text{I}}\right\rangle = 0
\end{equation}
\begin{equation}
2\left\langle\mathop{\tr} X_{\text{I}}X_{\text{I}}\right\rangle
+\left\langle\mathop{\tr} \Pi_{\text{I}}X_{\text{J}}X_{\text{I}}X_{\text{J}}\right\rangle
-\left\langle\mathop{\tr} X_{\text{I}}X_{\text{J}}X_{\text{I}}\Pi_{\text{J}}\right\rangle = 0
\end{equation}
\begin{equation}
\left\langle\mathop{\tr} \Pi_{\text{I}}X_{\text{J}}X_{\text{I}}X_{\text{J}}\right\rangle
+\left\langle\mathop{\tr} X_{\text{I}}X_{\text{J}}X_{\text{I}}\Pi_{\text{J}}\right\rangle = 0
\end{equation}
\begin{equation}
-\left\langle\mathop{\tr} X_{\text{I}}X_{\text{I}}\right\rangle
-\left\langle\mathop{\tr} X_{\text{I}}X_{\text{J}}\Pi_{\text{I}}X_{\text{J}}\right\rangle
+\left\langle\mathop{\tr} X_{\text{I}}X_{\text{J}}X_{\text{I}}\Pi_{\text{J}}\right\rangle = 0
\end{equation}
\begin{equation}
\left\langle\mathop{\tr} \Pi_{\text{I}}X_{\text{J}}X_{\text{I}}X_{\text{J}}\right\rangle
+\left\langle\mathop{\tr} X_{\text{I}}\Pi_{\text{J}}X_{\text{I}}X_{\text{J}}\right\rangle
+\left\langle\mathop{\tr} X_{\text{I}}X_{\text{J}}\Pi_{\text{I}}X_{\text{J}}\right\rangle
+\left\langle\mathop{\tr} X_{\text{I}}X_{\text{J}}X_{\text{I}}\Pi_{\text{J}}\right\rangle = 0
\end{equation}
\begin{equation}
\left\langle\mathop{\tr} X_{\text{I}}X_{\text{I}}\right\rangle
+\left\langle\mathop{\tr} \Pi_{\text{I}}X_{\text{I}}X_{\text{J}}X_{\text{J}}\right\rangle
-\left\langle\mathop{\tr} X_{\text{I}}X_{\text{J}}X_{\text{J}}\Pi_{\text{I}}\right\rangle
+D\left\langle\mathop{\tr} X_{\text{I}}X_{\text{I}}\right\rangle = 0
\end{equation}
\begin{equation}
\left\langle\mathop{\tr} \Pi_{\text{I}}X_{\text{J}}X_{\text{J}}X_{\text{I}}\right\rangle
+\left\langle\mathop{\tr} X_{\text{I}}X_{\text{J}}X_{\text{J}}\Pi_{\text{I}}\right\rangle = 0
\end{equation}
\begin{equation}
-\left\langle\mathop{\tr} X_{\text{I}}X_{\text{I}}\Pi_{\text{J}}X_{\text{J}}\right\rangle
+\left\langle\mathop{\tr} X_{\text{I}}X_{\text{J}}X_{\text{J}}\Pi_{\text{I}}\right\rangle
-D\left\langle\mathop{\tr} X_{\text{I}}X_{\text{I}}\right\rangle = 0
\end{equation}
\begin{equation}
\left\langle\mathop{\tr} \Pi_{\text{I}}X_{\text{J}}X_{\text{J}}X_{\text{I}}\right\rangle
+\left\langle\mathop{\tr} X_{\text{I}}\Pi_{\text{J}}X_{\text{J}}X_{\text{I}}\right\rangle
+\left\langle\mathop{\tr} X_{\text{I}}X_{\text{J}}\Pi_{\text{J}}X_{\text{I}}\right\rangle
+\left\langle\mathop{\tr} X_{\text{I}}X_{\text{J}}X_{\text{J}}\Pi_{\text{I}}\right\rangle = 0
\end{equation}
\begin{equation}
\left\langle\mathop{\tr} X_{\text{I}}X_{\text{I}}X_{\text{J}}X_{\text{J}}\right\rangle
-\left\langle\mathop{\tr} X_{\text{I}}X_{\text{J}}X_{\text{J}}X_{\text{I}}\right\rangle = 0
\end{equation}
\begin{equation}
\left\langle\mathop{\tr} X_{\text{I}}X_{\text{J}}X_{\text{J}}X_{\text{I}}\right\rangle
-\left\langle\mathop{\tr} X_{\text{I}}X_{\text{I}}X_{\text{J}}X_{\text{J}}\right\rangle = 0
\end{equation}
\begin{equation}
\left\langle\mathop{\tr} \Pi_{\text{I}}\Pi_{\text{I}}\right\rangle
+\left\langle\mathop{\tr} X_{\text{I}}X_{\text{I}}X_{\text{J}}X_{\text{J}}\right\rangle
-2\left\langle\mathop{\tr} X_{\text{I}}X_{\text{J}}X_{\text{I}}X_{\text{J}}\right\rangle
+\left\langle\mathop{\tr} X_{\text{I}}X_{\text{J}}X_{\text{J}}X_{\text{I}}\right\rangle
+M^2\left\langle\mathop{\tr} X_{\text{I}}X_{\text{I}}\right\rangle = 0
\end{equation}
\begin{equation}
    \begin{split}
        &\left\langle\mathop{\tr} \Pi_{\text{I}}X_{\text{I}}X_{\text{J}}X_{\text{J}}\right\rangle -2 \left\langle\mathop{\tr} \Pi_{\text{I}}X_{\text{J}}X_{\text{I}}X_{\text{J}}\right\rangle +\left\langle\mathop{\tr} \Pi_{\text{I}}X_{\text{J}}X_{\text{J}}X_{\text{I}}\right\rangle +\left\langle\mathop{\tr} X_{\text{I}}X_{\text{I}}X_{\text{J}}\Pi_{\text{J}}\right\rangle -2 \left\langle\mathop{\tr} X_{\text{I}}X_{\text{J}}X_{\text{I}}\Pi_{\text{J}}\right\rangle \\
        &+\left\langle\mathop{\tr} X_{\text{I}}X_{\text{J}}X_{\text{J}}\Pi_{\text{I}}\right\rangle +M^2\left\langle\mathop{\tr} \Pi_{\text{I}}X_{\text{I}}\right\rangle +M^2\left\langle\mathop{\tr} X_{\text{I}}\Pi_{\text{I}}\right\rangle =0.
    \end{split}
\end{equation}
\end{small}
We notice that, up to level 5, all the equations listed above are linear. Quadratic equations arise from the cyclicity properties of the moments, which introduce double-trace operators. The first quadratic equation appears at level 7, when we cyclically permute, for example, the moment $\left\langle\operatorname{tr} X_{\text{I}} X_{\text{J}} X_{\text{J}} \Pi_{\text{I}} X_{\text{K}} X_{\text{K}}\right\rangle$:
\begin{equation}
0 = \left\langle\operatorname{tr} X_{\text{I}} X_{\text{J}} X_{\text{J}} \Pi_{\text{I}} X_{\text{K}} X_{\text{K}}\right\rangle
- \left\langle\operatorname{tr} X_{\text{I}} X_{\text{I}} \Pi_{\text{J}} X_{\text{K}} X_{\text{K}} X_{\text{J}}\right\rangle
+ D \left\langle\operatorname{tr} X_{\text{I}} X_{\text{I}}\right\rangle^2
\end{equation}
Solving these equations at level $5$, we find that only three variables remain independent: $\left\langle\operatorname{tr} \Pi_{\text{I}} \Pi_{\text{I}} \right\rangle$, $\left\langle\operatorname{tr} X_{\text{I}} X_{\text{I}} \right\rangle$, and $\left\langle\operatorname{tr} X_{\text{I}} X_{\text{I}} X_{\text{J}} X_{\text{J}} \right\rangle$. The remaining variables are determined by:
\begin{equation}\label{eq: level5sol}
    \begin{split}
        &\left\langle\mathop{\tr} \Pi_{\text{I}}X_{\text{I}}\right\rangle \to -\frac{1}{2}D\\ &\left\langle\mathop{\tr} X_{\text{I}}\Pi_{\text{I}}\right\rangle \to \frac{1}{2}D\\ &\left\langle\mathop{\tr} X_{\text{I}}X_{\text{J}}X_{\text{I}}X_{\text{J}}\right\rangle \to \left\langle\mathop{\tr} X_{\text{I}}X_{\text{I}}X_{\text{J}}X_{\text{J}}\right\rangle +\frac{1}{2}M^2\left\langle\mathop{\tr} X_{\text{I}}X_{\text{I}}\right\rangle +\frac{1}{2} \left\langle\mathop{\tr} \Pi_{\text{I}}\Pi_{\text{I}}\right\rangle \\ &\left\langle\mathop{\tr} X_{\text{I}}X_{\text{J}}X_{\text{J}}X_{\text{I}}\right\rangle \to \left\langle\mathop{\tr} X_{\text{I}}X_{\text{I}}X_{\text{J}}X_{\text{J}}\right\rangle \\ &\left\langle\mathop{\tr} \Pi_{\text{I}}X_{\text{I}}X_{\text{J}}X_{\text{J}}\right\rangle \to -\frac{1}{2}D\left\langle\mathop{\tr} X_{\text{I}}X_{\text{I}}\right\rangle -\frac{1}{2} \left\langle\mathop{\tr} X_{\text{I}}X_{\text{I}}\right\rangle \\ &\left\langle\mathop{\tr} \Pi_{\text{I}}X_{\text{J}}X_{\text{I}}X_{\text{J}}\right\rangle \to -\left\langle\mathop{\tr} X_{\text{I}}X_{\text{I}}\right\rangle \\ &\left\langle\mathop{\tr} \Pi_{\text{I}}X_{\text{J}}X_{\text{J}}X_{\text{I}}\right\rangle \to -\frac{1}{2}D\left\langle\mathop{\tr} X_{\text{I}}X_{\text{I}}\right\rangle -\frac{1}{2} \left\langle\mathop{\tr} X_{\text{I}}X_{\text{I}}\right\rangle \\ &\left\langle\mathop{\tr} X_{\text{I}}\Pi_{\text{I}}X_{\text{J}}X_{\text{J}}\right\rangle \to \frac{1}{2}D\left\langle\mathop{\tr} X_{\text{I}}X_{\text{I}}\right\rangle -\frac{1}{2} \left\langle\mathop{\tr} X_{\text{I}}X_{\text{I}}\right\rangle \\ &\left\langle\mathop{\tr} X_{\text{I}}\Pi_{\text{J}}X_{\text{I}}X_{\text{J}}\right\rangle \to 0\\ &\left\langle\mathop{\tr} X_{\text{I}}\Pi_{\text{J}}X_{\text{J}}X_{\text{I}}\right\rangle \to -\frac{1}{2}D\left\langle\mathop{\tr} X_{\text{I}}X_{\text{I}}\right\rangle +\frac{1}{2} \left\langle\mathop{\tr} X_{\text{I}}X_{\text{I}}\right\rangle \\ &\left\langle\mathop{\tr} X_{\text{I}}X_{\text{I}}\Pi_{\text{J}}X_{\text{J}}\right\rangle \to -\frac{1}{2}D\left\langle\mathop{\tr} X_{\text{I}}X_{\text{I}}\right\rangle +\frac{1}{2} \left\langle\mathop{\tr} X_{\text{I}}X_{\text{I}}\right\rangle \\ &\left\langle\mathop{\tr} X_{\text{I}}X_{\text{I}}X_{\text{J}}\Pi_{\text{J}}\right\rangle \to \frac{1}{2}D\left\langle\mathop{\tr} X_{\text{I}}X_{\text{I}}\right\rangle +\frac{1}{2} \left\langle\mathop{\tr} X_{\text{I}}X_{\text{I}}\right\rangle \\ &\left\langle\mathop{\tr} X_{\text{I}}X_{\text{J}}\Pi_{\text{I}}X_{\text{J}}\right\rangle \to 0\\ &\left\langle\mathop{\tr} X_{\text{I}}X_{\text{J}}\Pi_{\text{J}}X_{\text{I}}\right\rangle \to \frac{1}{2}D\left\langle\mathop{\tr} X_{\text{I}}X_{\text{I}}\right\rangle -\frac{1}{2} \left\langle\mathop{\tr} X_{\text{I}}X_{\text{I}}\right\rangle \\ &\left\langle\mathop{\tr} X_{\text{I}}X_{\text{J}}X_{\text{I}}\Pi_{\text{J}}\right\rangle \to \left\langle\mathop{\tr} X_{\text{I}}X_{\text{I}}\right\rangle \\ &\left\langle\mathop{\tr} X_{\text{I}}X_{\text{J}}X_{\text{J}}\Pi_{\text{I}}\right\rangle \to \frac{1}{2}D\left\langle\mathop{\tr} X_{\text{I}}X_{\text{I}}\right\rangle +\frac{1}{2} \left\langle\mathop{\tr} X_{\text{I}}X_{\text{I}}\right\rangle 
    \end{split}
\end{equation}

\subsection{Positivity}

To express the positivity conditions, we classify all vectors in the $\mathrm{O}(D)$ irreducible representations up to level 2:
\begin{equation}\label{eq: veclevel5}
    \begin{split}
        &1, \quad X_{\text{I}} X_{\text{I}} \\
        &X_{\text{I}}, \quad \Pi_{\text{I}} \\
        &-\frac{1}{D} \delta_{\text{I}\text{J}} X_{\text{K}} X_{\text{K}} 
          + \frac{1}{2} X_{\text{I}} X_{\text{J}} + \frac{1}{2} X_{\text{J}} X_{\text{I}} \\
        &\frac{1}{2} X_{\text{I}} X_{\text{J}} - \frac{1}{2} X_{\text{J}} X_{\text{I}}
    \end{split}
\end{equation}
These correspond, respectively, to the singlet, vector, traceless-symmetric, and antisymmetric representations. When computing the inner product as defined in the main text, vectors belonging to different irreducible representations are orthogonal, as required by group theoretical considerations. The vectors in~\eqref{eq: veclevel5} should be regarded as tensors with abstract indices. The inner product is defined by first taking the trace of the product of matrices with their adjoints, followed by summing over the corresponding abstract indices from $1$ to $D$. As an example, the corresponding positivity condition for the vectors in~\eqref{eq: veclevel5} is given by:
\begin{equation}
    \begin{split}
    &\begin{pmatrix}
        1 & \left\langle\mathop{\tr} X_{\text{I}}X_{\text{I}}\right\rangle  \\
        \left\langle\mathop{\tr} X_{\text{I}}X_{\text{I}}\right\rangle  & \left\langle\mathop{\tr} X_{\text{I}}X_{\text{I}}X_{\text{J}}X_{\text{J}}\right\rangle  \\
    \end{pmatrix}\succeq 0,\quad 
    \begin{pmatrix}
        \left\langle\mathop{\tr} X_{\text{I}}X_{\text{I}}\right\rangle  & \left\langle\mathop{\tr} X_{\text{I}}\Pi_{\text{I}}\right\rangle  \\
        -\left\langle\mathop{\tr} \Pi_{\text{I}}X_{\text{I}}\right\rangle  & -\left\langle\mathop{\tr} \Pi_{\text{I}}\Pi_{\text{I}}\right\rangle  \\
    \end{pmatrix}\succeq 0, \\
    &-\frac{1}{D}\left\langle\mathop{\tr} X_{\text{I}}X_{\text{I}}X_{\text{J}}X_{\text{J}}\right\rangle +\frac{1}{2} \left\langle\mathop{\tr} X_{\text{I}}X_{\text{J}}X_{\text{I}}X_{\text{J}}\right\rangle +\frac{1}{2} \left\langle\mathop{\tr} X_{\text{I}}X_{\text{J}}X_{\text{J}}X_{\text{I}}\right\rangle\geq 0,\\
    &\frac{1}{2} \left\langle\mathop{\tr} X_{\text{I}}X_{\text{J}}X_{\text{J}}X_{\text{I}}\right \rangle -\frac{1}{2} \left\langle\mathop{\tr} X_{\text{I}}X_{\text{J}}X_{\text{I}}X_{\text{J}}\right \rangle\geq 0,  
\end{split}
\end{equation}
Substituting the solution from~\eqref{eq: level5sol}, we obtain the following inequality constraints for the three free variables:
\begin{equation}\label{eq: pos1}
    \begin{split}
       &\begin{pmatrix}
        1 & \left\langle\mathop{\tr} X_{\text{I}}X_{\text{I}}\right\rangle  \\
        \left\langle\mathop{\tr} X_{\text{I}}X_{\text{I}}\right\rangle  & \left\langle\mathop{\tr} X_{\text{I}}X_{\text{I}}X_{\text{J}}X_{\text{J}}\right\rangle  \\
       \end{pmatrix}\succeq 0,\quad \begin{pmatrix}
        \left\langle\mathop{\tr} X_{\text{I}}X_{\text{I}}\right\rangle  & \frac{1}{2}D \\
        \frac{1}{2}D & -\left\langle\mathop{\tr} \Pi_{\text{I}}\Pi_{\text{I}}\right\rangle  \\
       \end{pmatrix}\succeq 0, \\
        &\frac{1}{4} \left\langle\mathop{\tr} \Pi_{\text{I}}\Pi_{\text{I}}\right\rangle  +\frac{D-1}{D}\left\langle\mathop{\tr} X_{\text{I}}X_{\text{I}}X_{\text{J}}X_{\text{J}}\right\rangle +\frac{1}{4}M^2\left\langle\mathop{\tr} X_{\text{I}}X_{\text{I}}\right\rangle\geq 0\\
        & -\frac{1}{4}M^2\left\langle\mathop{\tr} X_{\text{I}}X_{\text{I}}\right\rangle -\frac{1}{4} \left\langle\mathop{\tr} \Pi_{\text{I}}\Pi_{\text{I}}\right\rangle\geq 0
    \end{split}
\end{equation}

Similarly, for the ground state positivity condition defined in the main text, we have:
\begin{equation}\label{eq: pos2}
    \begin{split}
       &\left\langle\operatorname{tr} X_{\text{I}} X_{\text{I}}\right\rangle \geq 0,\\[1ex]
       & \begin{pmatrix}
        \frac{1}{2} D & -\left\langle\operatorname{tr} \Pi_{\text{I}} \Pi_{\text{I}}\right\rangle  \\
        -\left\langle\operatorname{tr} \Pi_{\text{I}} \Pi_{\text{I}}\right\rangle  & (D-1)\left\langle\operatorname{tr} X_{\text{I}} X_{\text{I}}\right\rangle + \frac{1}{2} D M^2 \\
       \end{pmatrix} \succeq 0,\\[2ex]
       &\left( \frac{D+1}{2} - \frac{1}{D} \right)\left\langle\operatorname{tr} X_{\text{I}} X_{\text{I}}\right\rangle \geq 0, \\[1ex]
       &\frac{D-1}{2}\left\langle\operatorname{tr} X_{\text{I}} X_{\text{I}}\right\rangle \geq 0.
    \end{split}
\end{equation}

It turns out that the positivity conditions Eq.~\eqref{eq: pos1} and Eq.~\eqref{eq: pos2} carve out the allowed region for 
$\mathcal{E} \equiv \frac{E_0}{N^2} (g^2_\text{YM} N)^{-1/3}$ and $\ev{\tr X^2} = \frac{1}{D}\ev{\tr X_I X_I}$:
\begin{equation}\label{eq: analytic-supp}
\begin{split}
    \frac{\mathcal{E}}{D} &\ge \max\Big\{ \frac{3  }{16 \ev{\tr X^2} }+\frac{ M^2   \ev{\tr X^2}}{4},M^2 \ev{\tr X^2}\Big\},\\
    \frac{\mathcal{E}}{D} &\le 
  \frac{1}{8} \left[ 2 M^2  \ev{\tr X^2} + 3   \left(2 (D-1) \ev{\tr X^2}+M^2 \right)^{1/2}  \right]
\end{split}
\end{equation}

\subsection{Virial theorem derivation of the bounds}

As an illustration of the general bootstrap constraints and to identify the inequalities that are saturated, we present a simple derivation of Eq.~\eqref{eq: analytic-supp} based on the virial theorem.

Let us denote the kinetic, mass, and potential (commutator$^2$) contributions to the energy by
$
K,  \mathbf{M},  V,
$
respectively.  
The virial theorem implies
\[
-2K + 2\mathbf{M} + 4V = 0,
\]
while the total ground-state energy satisfies
\[
K + \mathbf{M} + V = E_0.
\]
Combining these relations, we obtain
\begin{align}
\mathcal{E}
\equiv
\frac{E_0}{N^2}(g^2_{\mathrm{YM}} N)^{-1/3}
=
\frac{3}{4}\,
\ev{\operatorname{tr} P_I P_I}
+
\frac{1}{4}\,
M^2\,
\ev{\operatorname{tr} X_I X_I}.
\end{align}

\paragraph{Inner-product positivity}
Applying inner-product positivity to the operators $\{X_I,\,-iP_I\}$, and also to the antisymmetric operator $X_I X_J - X_J X_I$, yields
\begin{equation}
\begin{pmatrix}
\ev{\operatorname{tr} X_I X_I} & D/2 \\
D/2 & \ev{\operatorname{tr} P_I P_I}
\end{pmatrix}
\succeq 0,
\qquad
\ev{\operatorname{tr} P_I P_I}
\ge
M^2\,
\ev{\operatorname{tr} X_I X_I}.
\end{equation}

\paragraph{Ground-state positivity}
Ground-state positivity applied to the operators $\{X_I, P_I\}$ gives
\begin{align}\label{eq: gsp2}
\begin{pmatrix}
\frac{D}{2} &
\ev{\operatorname{tr} P_I P_I}
\\[4pt]
\ev{\operatorname{tr} P_I P_I} &
(D-1)\ev{\operatorname{tr} X_I X_I}
+
\frac{D}{2}M^2
\end{pmatrix}
\succeq 0.
\end{align}

These constraints are sufficient to reproduce the analytic bound stated in Eq.~\eqref{eq: analytic-supp}.

\subsection{Higher levels}
The constraints discussed in this appendix so far are universal for any $D$. However, as we proceed to higher levels in the bootstrap, additional conditions arise that are non-universal and depend explicitly on the value of $D$. Thus, the role of $D$ in the bootstrap is not merely as a normalization factor, but also reflects deeper structural features of the theory.

One such example is the vanishing of certain $\delta$ products at $D=2$:
\begin{equation}
    0 = \varepsilon_{IJK} \varepsilon_{LMN} = \delta_{IL} \delta_{JM} \delta_{KN}
    - \delta_{IM} \delta_{JL} \delta_{KN}
    - \delta_{IL} \delta_{JN} \delta_{KM}
    - \delta_{IN} \delta_{JM} \delta_{KL}
    + \delta_{IM} \delta_{JN} \delta_{KL}
    + \delta_{IN} \delta_{JL} \delta_{KM}
\end{equation}
The right-hand side implies non-trivial kinematic relations among six-letter words, which explains why there are significantly fewer free variables in $D=2$ compared to $D=9$, as shown in Table I in the main text. A similar truncation occurs for the positivity conditions when considering the irreducible decomposition of operators. The first such example arises from the fact that there is no $(D+1)$-indexed antisymmetric irreducible representation in an $\mathrm{O}(D)$ symmetric matrix model, as guaranteed by the pigeonhole principle.

As an example of higher levels, we present the level 6 irreducible representation vectors corresponding to the following Young diagram (assuming $D \geq 3$):
\begin{equation}
    \; \begin{ytableau}
        1 & 2  \\
        3 \\
    \end{ytableau} \;
\end{equation}
\begin{small}
\begin{equation}
    \begin{split}
        &-\frac{1}{2}\frac{1}{D-1}\delta _{\text{I}\text{J}}   \mathop{\tr} X_{\text{L}}X_{\text{L}}X_{\text{K}}   +\frac{1}{4}\frac{1}{D-1}\delta _{\text{I}\text{J}}   \mathop{\tr} X_{\text{K}}X_{\text{L}}X_{\text{L}}   +\frac{1}{4}\frac{1}{D-1}\delta _{\text{I}\text{J}}   \mathop{\tr} X_{\text{L}}X_{\text{K}}X_{\text{L}}   \\& -\frac{1}{4}\frac{1}{D-1}\delta _{\text{J}\text{K}}   \mathop{\tr} X_{\text{I}}X_{\text{L}}X_{\text{L}}   -\frac{1}{4}\frac{1}{D-1}\delta _{\text{J}\text{K}}   \mathop{\tr} X_{\text{L}}X_{\text{I}}X_{\text{L}}   +\frac{1}{2}\frac{1}{D-1}\delta _{\text{J}\text{K}}   \mathop{\tr} X_{\text{L}}X_{\text{L}}X_{\text{I}}   \\&+\frac{1}{4}   \mathop{\tr} X_{\text{I}}X_{\text{J}}X_{\text{K}}   +\frac{1}{4}   \mathop{\tr} X_{\text{J}}X_{\text{I}}X_{\text{K}}   -\frac{1}{4}   \mathop{\tr} X_{\text{J}}X_{\text{K}}X_{\text{I}}   -\frac{1}{4}   \mathop{\tr} X_{\text{K}}X_{\text{J}}X_{\text{I}}  , \\ &-\frac{1}{2}\frac{1}{D-1}\delta _{\text{I}\text{J}}   \mathop{\tr} X_{\text{L}}X_{\text{K}}X_{\text{L}}   +\frac{1}{4}\frac{1}{D-1}\delta _{\text{I}\text{J}}   \mathop{\tr} X_{\text{K}}X_{\text{L}}X_{\text{L}}   +\frac{1}{4}\frac{1}{D-1}\delta _{\text{I}\text{J}}   \mathop{\tr} X_{\text{L}}X_{\text{L}}X_{\text{K}}   \\& -\frac{1}{4}\frac{1}{D-1}\delta _{\text{J}\text{K}}   \mathop{\tr} X_{\text{I}}X_{\text{L}}X_{\text{L}}   -\frac{1}{4}\frac{1}{D-1}\delta _{\text{J}\text{K}}   \mathop{\tr} X_{\text{L}}X_{\text{L}}X_{\text{I}}   +\frac{1}{2}\frac{1}{D-1}\delta _{\text{J}\text{K}}   \mathop{\tr} X_{\text{L}}X_{\text{I}}X_{\text{L}}   \\&+\frac{1}{4}   \mathop{\tr} X_{\text{I}}X_{\text{K}}X_{\text{J}}   -\frac{1}{4}   \mathop{\tr} X_{\text{J}}X_{\text{I}}X_{\text{K}}   +\frac{1}{4}   \mathop{\tr} X_{\text{J}}X_{\text{K}}X_{\text{I}}   -\frac{1}{4}   \mathop{\tr} X_{\text{K}}X_{\text{I}}X_{\text{J}}  ,
    \end{split}
\end{equation}
\end{small}

For $D=2$ the general irrep decomposition is in terms of charge. 
Let us consider an operator $\mathcal{O}_k $ with $U(1)$ charge $k$. It is related to an operator $\mathcal{O}_{I_1 I_2 \cdots I_k}$ in the fully symmetric representation via
\begin{align}
v^I &= (1,\i), \quad \delta_{IJ} v^I v^J = 0\\
    \mathcal{O}_k &= v^{I_1} v^{I_2} \cdots v^{I_k} \mathcal{O}_{I_1 \cdots I_k} .
\end{align}    
For example, let $\mathcal{O}_1$ and $\mathcal{O}_2$ be invariant under the O(2) transformations. Then the $k=2$ the operator $ \mathcal{O}_1 Z \mathcal{O}_2 Z $ may be written 
\begin{align}
 \mathcal{O}_1 Z \mathcal{O}_2 Z = v^I v^I \left[ \hf (\mathcal{O}_1 X_I  \mathcal{O}_2 X_J + \mathcal{O}_1 X_J \mathcal{O}_2 X_I) - \frac{1}{2} \delta_{IJ} \mathcal{O}_1 X_K \mathcal{O}_2 X_K \right].
\end{align}
Similarly, a charge $-k$ operator $\mathcal{O}_{-k}$ can be obtained in a similar fashion by using the vectors $\bar{v}^I = (1, -\i)$. 
Since the symmetry of the model is actually O(2) and not just U(1), the irreps are actually 2-dimensional for $k \ne 0$, spanned by $\{ \mathcal{O}_k, \mathcal{O}_{-k} \}$, with the reflection element $r \in $ O(2) acting by $U_{k}(r) \mathcal{O}_k =\mathcal{O}_{-k} $. So the irreps of O(2) may be labeled by just a non-negative integer $k \ge 0 $.

\appsection{Positivity and O(9) blocks}{app: irrep}
\def\RR{\mathcal{R}}
\def\O{\mathcal{O}}
\def\i{\mathbf{i}}
\def\j{\mathbf{j}}

Consider a collection of operators $O_{1, I_1, \cdots, I_n},O_{2, I_1, \cdots, I_n}, \cdots, O_{|\alpha|, I_1, \cdots, I_n} $ with vector indices. (Here we should think of $|\alpha|$ as the number of different ``words'' that have a level $\le$ to some cutoff.)  We denote their O$(D)$ indices by the shorthand  $\i = \{I_1, I_2, \cdots, I_n \}$. Positivity of the inner product yields %
\begin{equation}
    \mathcal{M}_{ \{\alpha, \i\} ,   \{ \beta, \j\} } =  \ev{ \tr \, \bar{O}_{\alpha,\i} O_{\beta, \j} }, \quad \mathcal{M} \succeq 0.
\end{equation}

If we write out this matrix explicitly, we will have a $|\alpha| D^{n} \times  |\alpha| D^n$ which will quickly become intractable for even modest $n$. %
Our goal instead is to use group theory to boil down all the positivity constraints of this explicit matrix into a much smaller set of positivity constraints on the O($D$) singlet operators. Since the group theory does not touch the $\alpha$ index, we will henceforth focus on a particular operator, say $\mathcal{O}_\i  = \mathcal{O}_{1, \i}$. 

To this end, we decompose the operators into irreps, e.g.,%
\begin{align}
  O_\i = \sum_R  \sum_{r=1}^{\dim R} (C_R)^{r}_\i (O_R)_r 
\end{align}
An irrep appears in the sum multiple times if the decomposition has multiplicity.
Then SO(9) invariance of the state implies that 
\begin{align} \label{mijBlock}
    \ev{\tr (\bar{O}_{\bar{R}})_{\bar{r}} (O_{R})_{r}} = \delta_{\bar{R},R} \delta_{\bar{r},r} a_{\bar{R},R},\\
        \mathcal{M}_{\i \j } = \sum_{R,\bar{R}, r} a_{\bar{R}, R} (C_{\bar R})_\i^r (C_R)_\j^r.
\end{align}
Here the symbol $\delta_{R,R'} = 1$ if $R$ and $R'$ are equivalent representations of SO(9), or else $\delta_{R,R'} = 0$. Thus we have parameterized a large matrix $\mathcal{M}$ in terms of a smaller number of coefficients $a_{\bar{R}, R}$. 
These coefficients are precisely just the O(9) singlet operators, e.g.,
\begin{align}
    a_{R,\bar{R}} = \ev{\tr (\bar{O}_{\bar{R}})_{r} (O_{R})_{r}}.
\end{align}
Furthermore, we can simplify the positivity requirement $\mathcal{M} \succeq 0$ by evaluating the requirement on a nice basis of vectors $\{ e_A\} $. %
A particularly nice basis is the following. First we view $(C_R)$ as a projector from the vector space indexed by $\mathcal{I}$ to the irrep $R$ (a smaller vector space indexed by $r$). Then we can define the basis $e_A$ to be a collection of the transposed projectors, where $A = (R_A, r_A)$. This spans the bigger vector space of dimension $|\mathcal{I}|$ and satisfies 
\begin{align}
  \sum_\j  e_{A}^{\j} (C_R)_\j^r = \delta_{R,R_A} \delta_{r_A}^r.
\end{align}
Then the positivity requirement can be expressed as %
\begin{align}
   \mathcal{M}_{AB} =  \bar{e}_{A}^{\i}  \mathcal{M}_{\i\j} e_{B}^{\j}  =  \sum_{R,\bar{R}} a_{\bar{R}_A, R_B} \delta_{\bar{R}_A,R_B}\delta_{\bar{r}_A, r_B} \succeq 0 . 
\end{align}
The conclusion is that we only need to impose
\begin{align}
    a_{\bar{R}, R } \succeq 0, \quad \bar{R} \sim R.
\end{align}
Note that even if the decomposition of $\i$ contains irrep $R$ with unit multiplicity, the generalization to multiple operators will typically lead to a non-trivial matrix $a$.

Let us consider a tensor of rank $n$. We are interested in the projectors 
\begin{align}
    (\pi_\mathsf{T})^{a_1 a_2 \cdots a_r}_{b_1 b_2 \cdots b_r}.  
\end{align}
These are defined to be orthonormal projectors:
\begin{align}
    (\pi_\mathsf{S})^{a_1 a_2 \cdots a_r}_{b_1 b_2 \cdots b_r}
    (\pi_\mathsf{T})^{b_1 b_2 \cdots b_r}_{c_1 c_2 \cdots c_r} =   \delta_{\mathsf{S} \mathsf{T}} \cdot  (\pi_\mathsf{T})^{a_1 a_2 \cdots a_r}_{c_1 c_2 \cdots c_r}  .
\end{align}
These projectors are also traceless, meaning
\begin{align}
 (\pi_\mathsf{T})^{a_1 a_2 \cdots a_r}_{b_1 b_2 \cdots b_r} \delta_{b_i b_j} = 0. \label{eq: tracelessCond}
\end{align}
A subtlety with this procedure occurs when a given irrep appears more than once in the above decomposition. 
In this case, the orthogonality of the projectors is potentially misleading. Since each degenerate irrep is isomorphic, there exists a linear map
\begin{align}
    \iota_{m,n}:  R_m \to R_n
\end{align}
Using this map, we can consider the contraction
\begin{align}
    (\pi_{R_m} ) ^{a_1 \cdots a_r}_{b_1 \cdots b_r}(\iota_{m,n})^{b_1 \cdots b_r}_{c_1 \cdots c_r}
     (\pi_{R_n} ) ^{c_1 \cdots c_r}_{a_1 \cdots a_r}(\iota_{1,m})
\end{align}
It will be convenient to simply consider the 
\begin{align}
    \iota_{m,n} = \iota_{m,1} \cdot \iota_{1,n}
\end{align}
Then, \eqref{mijBlock} becomes
\begin{align}
\mathcal{M}_{\i \j} &=   \mathcal{M}_{i_1 \cdots i_r}^{j_1 \cdots j_r} = \sum_{R} \sum_{m,n} a_{m,n}^R (\pi_{R_m})_{i_1 \cdots i_m}^{b_1 \cdots b_r} (\iota_{m,n})_{b_1 \cdots b_r}^{c_1 \cdots c_r} (\pi_{R_n})_{c_1 \cdots c_r}^{j_1 \cdots j_r}.\\
\mathcal{M}_{A B} &=  
(\pi_{R_m})^{i_1 \cdots i_m}_{c_1 \cdots c_r} \mathcal{M}_{i_1 \cdots i_m}^{j_1 \cdots j_m} (\pi_{R_n})_{j_1 \cdots j_m}^{b_1 \cdots b_r} (\iota_{n,m})_{b_1 \cdots b_r}^{c_1 \cdots c_r} 
\end{align}
The conclusion is that we can dress
\begin{align}
    \pi_m \to \pi_m \cdot \iota_{m, 1}
\end{align}

\subsection{Constructing the projectors}
\ytableausetup{centertableaux}
To construct the O($D$) projectors, we do so in a two step process, by first constructing the $GL(D)$ projectors, and then removing the traces to satisfy \eqref{eq: tracelessCond}.

For a given diagram, one can define the $GL(D)$ projector by first symmetrizing over the rows and then anti-symmetrizing over the columns of the given diagram:
\begin{align}
    \tilde{\pi}_{\text{Young Diagram}} = \prod_{\text{columns}} \text{antisymmetrizer} \cdot  \prod_{\text{rows}} \text{symmetrizer}
\end{align}

To illustrate this procedure concretely, let us consider the mixed-symmetry {\bf 231} irreps. This appears in the rank 3 tensor decomposition:
\begin{align}
   {\bf 9} \times \mathbf{9} \times \mathbf{9} = 3 (\mathbf{9}) + \mathbf{84} + \mathbf{156} + 2 (\mathbf{231} ). \end{align}
There are two isomorphic {\bf 231} irreps, corresponding to the two Young diagrams

\begin{align}
{\bf 231_+ } \; &=\; \begin{ytableau}
      1 & 2  \\
 3\\ \end{ytableau} \; = \; 
 \begin{tikzpicture}[baseline={([yshift=-.55ex]current bounding box.center)}, scale=1, every node/.style={inner sep=0}]
\draw (-1.5,0.2) -- (-.75,0.2);
\draw (-1.5,-0.2) -- (-.75,-0.2);
\draw (-1.5,-.6) -- (0.6,-.6);
\node[draw, fill=white, minimum width=0.3cm, minimum height=0.8cm] (W1) at (-0.75,0) {};
\node[draw, fill=black, minimum width=0.3cm, minimum height=0.8cm] (B1) at (0,-0.4) {};
\draw (-.6,0.2) -- (.6,0.2);
\draw (-.6,-0.2) -- (.6,-0.2);
\end{tikzpicture} ,\\
{\bf 231_-} \; &= \; \begin{ytableau}
      1 & 3  \\
 2\\ \end{ytableau} \; = \; 
\begin{tikzpicture}[baseline={([yshift=-.6ex]current bounding box.center)}, scale=1, every node/.style={inner sep=0}]
\draw (-2,0.2) -- (-.75,0.2) ;
\draw[black] (-2,-0.2) -- (-1.5,-0.2) -- (-1,-.6) -- (-.4,-.6) -- (.1,-.2) -- (1,-.2) -- (1.5,-.6) -- (1.7,-.6);
\draw[black] (-2,-.6) -- (-1.5,-.6)  -- (-1,-.2) -- (-.8,-.2); %
\node[draw, fill=white, minimum width=0.3cm, minimum height=0.8cm] (W1) at (-0.75,0) {};
\node[draw, fill=black, minimum width=0.3cm, minimum height=0.8cm] (B1) at (.5,0) {};
\draw (-.6,0.2) -- (1.7,0.2) ;
\draw[black] (-.6,-0.2) -- (-.4,-.2) -- (.1,-.6) --(1,-.6) -- (1.5, -.2) -- (1.7,-.2);
\end{tikzpicture} 
\end{align}
Here we are using ``birdtracks'' notation; we have indicated the symmetrizer using a white box and an anti-symmetrizer using a black box \cite{Cvitanovic:2008zz}. This defines some projectors $\tilde{\pi}^{a_1 \cdots a_r}_{b_1 \cdots b_r}$ which satisfy 
\begin{align} \label{eq: symm}
    \tilde{\pi}^{a_1 \cdots  a_r}_{b_1 \cdots b_r} -   \tilde{\pi}^{\tau( a_1 \cdots a_r)}_{b_1 \cdots b_r} = 0, \quad &\tau \in \{\text{row transpositions} \} \\ \label{eq: antisymm}
     \tilde{\pi}^{a_1 \cdots  a_r}_{b_1 \cdots b_r} +   \tilde{\pi}^{ a_1 \cdots a_r}_{\tau(b_1 \cdots b_r)} = 0, \quad &\tau \in \{\text{column transpositions} \} 
\end{align}
Here row (column) transpositions generate the subgroup of the permutation group that preserve the row (column) structure of the Young tableau.
Now to obtain the O($9$) projectors, we also need to subtract off any lower irreps (in this case $\mathbf{9}$). To do so we construct various Wick contractions, e.g.,
\begin{align} \label{delta1}
\delta_{a_1 a_2} \delta^{b_1 b_2} \delta_{a_3}^{b_3} \quad  = \quad \begin{tikzpicture}[baseline={([yshift=-.55ex]current bounding box.center)}, scale=1, every node/.style={inner sep=0}]
  \draw (1,1) -- (.5,1) arc[start angle=90,end angle=270,radius=.3cm] -- (1,.4);
  \draw (-1,1) -- (-0.5,1) arc[start angle=90,end angle=-90,radius=.3cm] -- (-1,.4);
\draw (-1,0) -- (1,0);
\end{tikzpicture}, \quad \delta_{a_2 a_3} \delta^{b_2 b_3} \delta_{a_1}^{b_1} \quad = \quad 
\begin{tikzpicture}[baseline={([yshift=-.55ex]current bounding box.center)}, scale=1, every node/.style={inner sep=0}]
    \draw (-1,1.4) -- (1,1.4);
  \draw (1,1) -- (.5,1) arc[start angle=90,end angle=270,radius=.3cm] -- (1,.4);
  \draw (-1,1) -- (-0.5,1) arc[start angle=90,end angle=-90,radius=.3cm] -- (-1,.4);
\end{tikzpicture},
\end{align}
\begin{align}\label{delta2}
\delta_{a_1  a_3} \delta^{b_1 b_3} \delta_{a_2}^{b_2} \quad = \quad  
\begin{tikzpicture}[baseline={([yshift=-.55ex]current bounding box.center)}, scale=1, every node/.style={inner sep=0}]
    \draw (-1,.5) -- (1,.5);
  \draw (1,1) -- (.75,1) arc[start angle=90,end angle=270,radius=.5cm] -- (1,0);
  \draw (-1,1) -- (-0.75,1) arc[start angle=90,end angle=-90,radius=.5cm] -- (-1,0);
\end{tikzpicture},
\end{align}
Then for each irrep/Young diagram $Y$, we sandwich these Wick contractions according to $\tilde\Delta_Y =  \tilde\pi \cdot \Delta \cdot \tilde\pi$. This results in a tensor with the same symmetry properties as \eqref{eq: symm} and \eqref{eq: antisymm}. For example, taking the first tensor in \eqref{delta1} and using the  $\mathbf{231}_+$ projector gives 
\begin{align}
     \tilde{\Delta}_{\mathbf{231}_+} \quad = \quad \begin{tikzpicture}[baseline={([yshift=-.55ex]current bounding box.center)}, scale=1, every node/.style={inner sep=0}]
\draw (-1.5,-.6) -- (2,-.6);
\draw (2,0.2) -- (1.35,0.2);
\draw (2,-0.2) -- (1.5,-0.2);
\draw (-1.5,0.2) to[out=0,in=90] (-0.1,0) to[out=-90,in=0] (-1.5,-0.2);
\draw (1.35,0.2) to[out=180,in=90] (0.1,0) to[out=-90,in=180] (1.35,-0.2);
\node[draw, fill=white, minimum width=0.3cm, minimum height=0.8cm] (W1) at (-1,0) {};
\node[draw, fill=white, minimum width=0.3cm, minimum height=0.8cm] (W2) at (1.0,0) {};
\node[draw, fill=black, minimum width=0.3cm, minimum height=0.8cm] (B1) at (-0.5,-0.4) {};
\node[draw, fill=black, minimum width=0.3cm, minimum height=0.8cm] (B2) at (1.5,-0.4) {};
\end{tikzpicture}   
\end{align}
Sandwiching the second tensor in \eqref{delta1} gives zero, and sandwiching \eqref{delta2} produces a tensor proportional to the above. 
We introduce the $O(D)$ inner product on tensors 
\begin{align}
\ev{S,T} &= \delta_{i_1 k_1} \cdots \delta_{i_r k_r} \delta^{j_1 l_1} \cdots \delta^{j_s l_s} S^{i_1 \cdots i_r}_{j_1 \cdots j_s} T^{k_1 \cdots k_r}_{l_1 \cdots l_s}\\
&=S_{k_1 \cdots k_r}^{l_1 \cdots l_s} T^{k_1 \cdots k_r}_{l_1 \cdots l_s}.
\end{align}
In general, we should carry out a Gram-Schmidt procedure using this inner product on the resulting set $\{\tilde\Delta_Y\} \xrightarrow{\text{GS}} \{ \Delta_Y\}$; in this example this produces only one tensor. We may choose the normalization of $\Delta_Y$ so that $(\Delta_Y)_i \cdot (\Delta_Y)_j = \delta_{ij} (\Delta_Y)_i$
We then demand that the final projector $\pi_Y$ is orthogonal to $\{ \Delta_Y \}$:
\begin{align}
&\ev{\pi_Y, \Delta_Y} = 0\\
&\pi_{{\bf 231_+}} \quad = \quad \begin{tikzpicture}[baseline={([yshift=-.55ex]current bounding box.center)}, scale=1, every node/.style={inner sep=0}]
\draw (-1.5,0.2) -- (-.75,0.2);
\draw (-1.5,-0.2) -- (-.75,-0.2);
\draw (-1.5,-.6) -- (0.6,-.6);
\node[draw, fill=white, minimum width=0.3cm, minimum height=0.8cm] (W1) at (-0.75,0) {};
\node[draw, fill=black, minimum width=0.3cm, minimum height=0.8cm] (B1) at (0,-0.4) {};
\draw (-.6,0.2) -- (.6,0.2);
\draw (-.6,-0.2) -- (.6,-0.2);
\end{tikzpicture} \, - \,
 \begin{tikzpicture}[baseline={([yshift=-.55ex]current bounding box.center)}, scale=1, every node/.style={inner sep=0}]
\draw (-1.5,-.6) -- (2,-.6);
\draw (2,0.2) -- (1.35,0.2);
\draw (2,-0.2) -- (1.5,-0.2);
\draw (-1.5,0.2) to[out=0,in=90] (-0.1,0) to[out=-90,in=0] (-1.5,-0.2);
\draw (1.35,0.2) to[out=180,in=90] (0.1,0) to[out=-90,in=180] (1.35,-0.2);
\node[draw, fill=white, minimum width=0.3cm, minimum height=0.8cm] (W1) at (-1,0) {};
\node[draw, fill=white, minimum width=0.3cm, minimum height=0.8cm] (W2) at (1.0,0) {};
\node[draw, fill=black, minimum width=0.3cm, minimum height=0.8cm] (B1) at (-0.5,-0.4) {};
\node[draw, fill=black, minimum width=0.3cm, minimum height=0.8cm] (B2) at (1.5,-0.4) {};
\end{tikzpicture}   
\end{align}

As another example, consider the mixed symmetry irrep
\begin{align}
    \begin{ytableau}
      1 & 2  \\
 3 & 4 \\ \end{ytableau}  \; = \; 
 \begin{tikzpicture}[baseline={([yshift=-.55ex]current bounding box.center)}, scale=1, every node/.style={inner sep=0}]
\draw (-1.5,0.2) -- (1,0.2);
\draw (-1.5,-0.2) -- (-.4,-0.2) -- (0,-.6) -- (1,-.6) ; %
\draw (-1.5,-.6) -- (-.4,-.6) -- (0,-0.2) -- (1,-.2) ; %
\draw (-1.5,-1) -- (1,-1);
\node[draw, fill=white, minimum width=0.3cm, minimum height=0.7cm] (W1) at (-0.75,0) {};
\node[draw, fill=white, minimum width=0.3cm, minimum height=0.7cm] (W1) at (-0.75,-0.8) {};
\node[draw, fill=black, minimum width=0.3cm, minimum height=0.7cm] (B1) at (0.5,0) {};
\node[draw, fill=black, minimum width=0.3cm, minimum height=0.7cm] (B1) at (0.5,-0.8) {};
\end{tikzpicture}
\end{align}
 The orthonormalization procedure on the Wick contractions produces 2 vectors $\{\Delta_Y\}$  and we find $\pi_Y = \tilde{\pi}_Y - (\Delta_Y)_1 - (\Delta_Y)_2$.

\appsection{Large \texorpdfstring{$D$}{D} expansion}{app: largeD}
\def\d{\mathrm{d}}
\def\tl{\tilde{\lambda}}
\renewcommand{\i}{\mathrm{i}}

In this appendix, we review the solution to the large $N$ and large $D$ Yang-Mills quantum mechanics and compute the 4-pt function. In this limit, we take $N\to \infty, D\to\infty$ holding fixed $\tl = g^2_\ym N D$. (More precisely, we hold fixed the dimensionless combination $\tl \, \tau^3$ where $\tau$ is some timescale in the problem.)

We will use a diagrammatic approach, see \cite{Hotta:1998en, Mandal:2009vz, Anegawa:2024qoz}. The bare propagator %
\begin{align}
\ev{(X_I)_{ij} (X_J)_{kl}}_\text{bare} \quad = \quad   \begin{tikzpicture}[>=stealth,baseline=(current bounding box.center)]
 \draw[ydouble](-2,0)--(0,0);
\end{tikzpicture} \quad = \quad \frac{1}{(\omega-(\alpha_j- \alpha_i))^2  } \delta_{il} \delta_{jk} \delta_{IJ}
\end{align}
Here $\alpha_i$ are the eigenvalues of the gauge field $A_0$.
We also have the two vertices from the potential:
\begin{align}
    &V= \frac{g^2_\ym}{4} \sum_{I,J} 2 \, {\color{blue} \Tr X_I^2 X_J^2} - 2 \, {\color{red} \Tr (X_I X_J)^2} \\
&\begin{tikzpicture}[>=stealth,baseline=(current bounding box.center)]
\node at (-1.2,1.2) {$I$};
\node at (1.2,1.2) {$I$};
\node at (1.2,-1.2) {$J$};
\node at (-1.2,-1.2) {$J$};
\draw[ydouble](-1,1)--(0.2,-0.2);
\draw[ydouble](-1,-1)--(0.2,0.2);
\draw[ydouble](0,0)--(1,1);
\draw[ydouble](0,0)--(1,-1);
\fill[blue] (0,0) circle (0.1);
\end{tikzpicture} \quad  =   -2 g^2_\ym   \\
&\begin{tikzpicture}[>=stealth,baseline=(current bounding box.center)]
\node at (-1.2,1.2) {$I$};
\node at (1.2,1.2) {$J$};
\node at (1.2,-1.2) {$I$};
\node at (-1.2,-1.2) {$J$};
\draw[ydouble](-1,1)--(0.2,-0.2);
\draw[ydouble](-1,-1)--(0.2,0.2);
\draw[ydouble](0,0)--(1,1);
\draw[ydouble](0,0)--(1,-1);
\fill[red] (0,0) circle (0.1);
\end{tikzpicture} \quad =  2 g^2_\ym  
\end{align}

At large $N$ and large $D$, the diagrams that contribute are the planar bubble diagrams. The most important vertex is the blue vertex; at large $N$ bubble diagrams that are made of red vertices are either non-planar or are suppressed by $1/D$ (due to the index contraction pattern). %

We can sum bubble diagrams using the Schwinger-Dyson equation:
\begin{align}
\begin{tikzpicture}[>=stealth,baseline={([yshift=-.5ex] current bounding box.center)}]
 \draw[Xdouble](-2,0)--(0,0);
  \node at (-2.25,0) {$I$};
 \node at (0.25,0) {$I$};
\end{tikzpicture} \quad &= \quad 
\begin{tikzpicture}[>=stealth,baseline={([yshift=-.5ex] current bounding box.center)}]
 \draw[ydouble](-2,0)--(0,0);
\end{tikzpicture} \quad +\quad 
\begin{tikzpicture}[>=stealth,baseline={([yshift=-2.5ex] current bounding box.center)}]
 \draw[ydouble](-2,0)--(0,0);
\draw[Xdouble] (0,0.5) circle (0.5);
\draw[Xdouble](0,0)--(1.5,0);
\node at (0,.5) {$K$};
\node at (1.75,0) {$I$};
\node at (-2.25,0) {$I$};
\fill[blue] (0,0) circle (0.1);
\end{tikzpicture}\\
G(\omega)& = \frac{1}{(\omega-(\alpha_j-\alpha_i))^2}-\frac{1}{\omega^2}\Delta^2 G(\omega), \quad \Delta^2 = 2g^2_\ym D N \int \frac{\d \omega'}{2\pi} G(\omega')  \\
G(\omega)& =\frac{1}{(\omega-(\alpha_j-\alpha_i))^2 + \Delta^2}, \quad \Delta^2 = \frac{g^2_\ym D N}{\Delta} \Rightarrow \Delta = \tilde{\lambda}^{1/3}
\end{align}
Here we have adopted the shorthand for the dressed propagator $G(\omega) = G_{ijji}(\omega)$ and $G(\tau)$ is the Fourier transform:
\begin{align}
	G_{ijkl}(\tau) = \frac{1}{2 \Delta_0  } e^{\i (\alpha_j -\alpha_i) |\tau| -\Delta |\tau| } \delta_{il}\delta_{jk}  \RA \ev{\tr X_I X_I} = \frac{D}{2\tl^{1/3}} 
\end{align}
This agrees with the leading answer in $1/D$ quoted in the main text and derived in \cite{Mandal:2009vz}. 
Note that in the final answer, the eigenvalues of the gauge field drop out. In the confining phase, we can essentially neglect the eigenvalues. Since the ground state is an SU($N$) singlet in the ungauged model, we should get identical correlators at zero temperature if we study the ungauged model where there is no gauge field.

\subsection{Four-point function}
We now compute the 4-pt function at equal times.
It is easier to compute the alternating correlator. This is given by dressing the bare interaction vertex:
\begin{align}
\ev{	\tr X_I X_J X_I X_J}_\text{c} = 
\begin{tikzpicture}[>=stealth,baseline=(current bounding box.center)]
\draw[Xdouble](-1,1)--(0,0);
\draw[Xdouble](-1,-1)--(0,0);
\draw[Xdouble](-.2,-.2)--(1,1);
\draw[Xdouble](-.2,.2)--(1,-1);
\fill[red] (0,0) circle (0.1);
\end{tikzpicture}
=  2 D^2 \frac{\tilde{g}^2 N}{D} \int_{-\infty}^\infty \d \tau \,  G(\tau)^4
= \frac{D \tilde{\lambda }}{16 \Delta^5} = \frac{D }{16
\tilde\lambda^{2/3}}%
\end{align}
Here we have used
\begin{align}
	\int_{-\infty} ^\infty \d \tau \,  G(\tau)^4 = \frac{2}{64 \Delta_0^5 }. %
\end{align}
Since the $O(D)$ indices are not aligned, the disconnected component is down by a factor of $1/D$, but contributes at the same order as the connected diagram:
\begin{align}
	\ev{	\tr X_I X_J X_I X_J} &= 
\begin{tikzpicture}[>=stealth,baseline=(current bounding box.center)]
\node at (-1.2,1.2) {$I$};
\node at (-1.2,-1.2) {$J$};
\node at (1.2,1.2) {$J$};
\node at (1.2,-1.2) {$I$};
\draw[Xdouble](-1,1)--(0,0);
\draw[Xdouble](-1,-1)--(0,0);
\draw[Xdouble](-.2,-.2)--(1,1);
\draw[Xdouble](-.2,.2)--(1,-1);
\fill[red] (0,0) circle (0.1);
\end{tikzpicture}
\quad + \quad 
\begin{tikzpicture}[>=stealth,baseline=(current bounding box.center)]
\node at (-.6,1.2) {$I$};
\node at (-.6,-1.2) {$J$};
\node at (.6,1.2) {$J$};
\node at (.6,-1.2) {$I$};
    \draw[Xdouble] (-.6,.8)--(-.6,-.8);
    \draw[Xdouble] (.6,.8)--(.6,-.8);
\end{tikzpicture}
\quad + \quad 
\begin{tikzpicture}[>=stealth,baseline=(current bounding box.center)]
\node at (-1.2,.6) {$I$};
\node at (-1.2,-.6) {$J$};
\node at (1.2,.6) {$J$};
\node at (1.2,-.6) {$I$};
    \draw[Xdouble] (-1,.6)--(1,.6);
    \draw[Xdouble] (-1,-.6)--(1,-.6);
\end{tikzpicture}
\\
&=\frac{D}{4\tilde{\lambda}^{2/3} }  +   \frac{D}{4\tilde{\lambda}^{2/3} } +\frac{D}{16 \tilde{\lambda}^{2/3} }
	= \frac{9 D^{1/3} }{16 \lambda^{2/3} }
\end{align}
One can then use the Virial theorem $-2K + 4V =0$ together with $K+V = \mathcal{E}$ to convert this to give the other level 4 correlator
\begin{align}
    \frac{2\mathcal{E}}{3}  = \frac{\tl^{2/3}}{D^{2/3}} \sum_{I,J}  \ev{\tr X_I^2 X_J^2}- \ev{\tr X_I X_J X_I X_J} .
\end{align}
As a result, we obtain the result cited in the main text:
\begin{align}
  \ev{\tr X_I^2 X_J^2} &=   \frac{D^2}{\tl^{2/3} }  \left [\qrt  + \frac{1}{D} \left( \frac{\sqrt{5}}{3}-\frac{9}{32} \right)  + \cdots\right].  \label{largeD3A}
\end{align}

\subsection{Non-alternating correlator}
As a check of the above computation, we can also compute the non-alternating correlator directly. In the O($D$) vector model context, this correlator is a ``double trace'' in the O($D$) indices; the leading term is given by the factorized answer. To compute the first non-trivial correction, we consider the connected Feynman diagrams:
\begin{align}
\ev{	\tr X_I X_I X_J X_J}_\text{c} &= 
\begin{tikzpicture}[>=stealth, scale=0.8, every node/.style={scale=0.8},baseline=(current bounding box.center)]
\node at (-1.2,1.2) {$I$};
\node at (-1.2,-1.2) {$I$};
\node at (1.2,1.2) {$J$};
\node at (1.2,-1.2) {$J$};
\draw[Xdouble](-1,1)--(0,0);
\draw[Xdouble](-1,-1)--(0,0);
\draw[Xdouble](-.2,-.2)--(1,1);
\draw[Xdouble](-.2,.2)--(1,-1);
\fill[blue] (0,0) circle (0.1);
\end{tikzpicture}
 + 
\begin{tikzpicture}[>=stealth, scale=0.8, every node/.style={scale=0.8},baseline=(current bounding box.center)]
\node at (-1.2,1.2) {$I$};
\node at (-1.2,-1.2) {$I$};
\node at (2.7,1.2) {$J$};
\node at (2.7,-1.2) {$J$};
\node at (.75,.75) {$K$};
\node at (.75,-.75) {$K$};
    \draw[Xdouble] (-1,1) -- (0,0);
    \draw[Xdouble] (-1,-1) -- (0,0);
    \draw[Xdouble] (0,0) to[out=30, in=150] (1.5,0);
    \draw[Xdouble] (0,0) to[out=-30, in=210] (1.5,0);
    \draw[Xdouble] (1.5,0) -- (2.5,1);
    \draw[Xdouble] (1.5,0) -- (2.5,-1);
    \fill[blue] (0,0) circle (0.1);
    \fill[blue] (1.5,0) circle (0.1);
\end{tikzpicture}  + 
\begin{tikzpicture}[>=stealth, scale=.8, every node/.style={scale=.8},baseline=(current bounding box.center)]
\node at (-1.2,1.2) {$I$};
\node at (-1.2,-1.2) {$I$};
\node at (4.2,1.2) {$J$};
\node at (4.2,-1.2) {$J$};
 \node at (.75,.75) {$K$};
\node at (.75,-.75) {$K$};
 \node at (2.25,.75) {$L$};
\node at (2.25,-.75) {$L$};
    \draw[Xdouble] (-1,1) -- (0,0);
    \draw[Xdouble] (-1,-1) -- (0,0);
    \draw[Xdouble] (0,0) to[out=30, in=150] (1.5,0);
    \draw[Xdouble] (0,0) to[out=-30, in=210] (1.5,0);
    \draw[Xdouble] (1.5,0) to[out=30, in=150] (3,0);
    \draw[Xdouble] (1.5,0) to[out=-30, in=210] (3,0);
    \draw[Xdouble] (3,0) -- (4,1);
    \draw[Xdouble] (3,0) -- (4,-1);
    \fill[blue] (0,0) circle (0.1);
    \fill[blue] (1.5,0) circle (0.1);
    \fill[blue] (3,0) circle (0.1);
\end{tikzpicture}  +  \cdots \\
&= - {\tl D} \frac{1}{\omega_1^2 + \Delta^2}\frac{1}{\omega_2^2 + \Delta^2}\frac{1}{\omega_3^2 + \Delta^2}\frac{1}{\omega_4^2 + \Delta^2}(1  - \tilde{\lambda} B(\omega_{12}))
\end{align}
Note the symmetry factors of $1/2$. Then one can sum over such bubble diagrams \cite{Anegawa:2024qoz} which produces the effective propagator (in the large $D$ vector model context, this is the $\sigma$ propagator):
\begin{align}
    B(\omega)  = \; \begin{tikzpicture}[>=stealth,baseline={([yshift=-.5ex] current bounding box.center)}]   \draw[bprop] (0,0)--(1.5,0);   \end{tikzpicture} \; =  \frac{1}{\tilde\lambda^{1/3}  } \frac{1}{\omega^2 + 5  \tl^{2/3} } , \quad B(\tau) = \frac{1}{2 \sqrt{5} \tl^{1/3} \tl^{1/3}}e^{-\sqrt{5} \tl^{1/3} |\tau|}
\end{align}
So to compute the bubble diagram correction we have
\begin{align}
    \int \d \tau_1 \d \tau_2 \, G^2(\tau_1) B(\tau_2 - \tau_1) G^2(-\tau_2) &= \frac{1}{2^5 \tl^{6/3}} \int \frac{\d \tilde\tau_1 \d \tilde\tau_2}{\sqrt{5} \tl^{2/3}}   \exp \left[- (2 |\tilde\tau_1| + 2  |\tilde\tau_2| + \sqrt{5} |\tilde\tau_2 - \tilde\tau_1| ) \right]\\
    &= \frac{16-7 \sqrt{5}}{32 \sqrt{5}  \tl^{8/3} }
\end{align}
We also need the 1-loop corrected disconnected contribution
\begin{align}
    \ev{\tr X_I X_I}  &=  \frac{D}{2\tl^{1/3}} \left[ 1 + \frac{2}{D} \left(\frac{7 \sqrt{5}}{30} - \frac{9}{32}\right)  + \cdots  \right] 
\end{align}
Putting together these ingredients,
\begin{align}
\ev{	\tr X_I X_I X_J X_J} &= \begin{tikzpicture}[>=stealth,baseline=(current bounding box.center)]
\node at (-1.2,1.2) {$I$};
\node at (-1.2,-1.2) {$I$};
\node at (1.2,1.2) {$J$};
\node at (1.2,-1.2) {$J$};
\draw[Xdouble](-1,1)--(0,0);
\draw[Xdouble](-1,-1)--(0,0);
\draw[Xdouble](-.2,-.2)--(1,1);
\draw[Xdouble](-.2,.2)--(1,-1);
\fill[blue] (0,0) circle (0.1);
\end{tikzpicture} \; + \;
\begin{tikzpicture}[>=stealth,baseline=(current bounding box.center)]
\draw[Xdouble](-1,1)--(0,0);
\draw[Xdouble](-1,-1)--(0,0);
\draw[bprop] (0,0)--(0.95,0);
\draw[Xdouble](0.95,0)--(1.95,1);
\draw[Xdouble](0.95,0)--(1.95,-1);
\fill[blue] (0,0) circle (0.1);
\fill[blue] (0.95,0) circle (0.1);
\end{tikzpicture}
\; + \;
 \begin{tikzpicture}[>=stealth,baseline=(current bounding box.center)]
\node at (-.6,1.2) {$I$};
\node at (-.6,-1.2) {$I$};
\node at (.6,1.2) {$J$};
\node at (.6,-1.2) {$J$};
    \draw[Xdouble] (-.6,.8)--(-.6,-.8);
    \draw[Xdouble] (.6,.8)--(.6,-.8);
\end{tikzpicture}
\; + \;
\begin{tikzpicture}[>=stealth,baseline=(current bounding box.center)]
\node at (-1.2,.6) {$I$};
\node at (-1.2,-.6) {$I$};
\node at (1.2,.6) {$J$};
\node at (1.2,-.6) {$J$};
    \draw[Xdouble] (-1,.6)--(1,.6);
    \draw[Xdouble] (-1,-.6)--(1,-.6);
\end{tikzpicture}\\
&= -\hf \frac{D}{16 \tl^{2/3}} +  \tl^2 \frac{16-7 \sqrt{5}}{32 \sqrt{5}  \tl^{8/3} } + \frac{D^2}{4 \tl^{2/3}}\left[1 + \frac{4}{D}\left(\frac{7 \sqrt{5}}{30} - \frac{9}{32}\right)\right] + \frac{D}{4 \tl^{2/3}} + \cdots \\
&=\frac{D^2}{ \tl^{2/3}}\left[ \frac{1}{4} + \frac{1}{D}\left(\frac{\sqrt5}{3} -\frac{9}{32} \right) + \cdots  \right]  %
\end{align}
We again recover the answer quoted in the main text.

\bibliography{Bosref}

\end{document}